\documentclass[useAMS]{mn2e}

\usepackage{graphicx}
\usepackage{color}

\title[Fundamental stellar parameters and age--metallicity relation of {\it Kepler} red giants]
{Fundamental stellar parameters and age--metallicity relation of {\it Kepler} red giants 
in comparison with theoretical evolutionary tracks
}

\author[Y. Takeda, et al.]
{Y. Takeda$^{1,2}$\thanks{E-mail:
takeda.yoichi@nao.ac.jp}\footnotemark[0]\thanks{
Based on data collected at Subaru Telescope, which is operated by the 
National Astronomical Observatory of Japan.}, 
A. Tajitsu$^{3}$, B. Sato$^{4}$, Y.-J. Liu$^{5}$, Y.-Q. Chen$^{5}$, and G. Zhao$^{5}$
\\
$^{1}$National Astronomical Observatory of Japan, 
2-21-1 Osawa, Mitaka, Tokyo 181-8588, Japan\\
$^{2}$The Graduate University for Advanced Studies (SOKENDAI), 2-21-1 Osawa, Mitaka, Tokyo 181-8588, Japan\\
$^{3}$Subaru Telescope, 650 N. A'ohoku Place, Hilo, HI 96720, U.S.A.\\
$^{4}$Department of Earth and Planetary Sciences, Tokyo Institute of Technology, 
2-12-1 Ookayama, Meguro-ku, Tokyo 152-8551, Japan\\
$^{5}$National Astronomical Observatories, Chinese Academy of Sciences,
A20 Datun Road, Chaoyang, Beijing 100012, P. R. China\\
}
\begin{document}

\date{Accepted 2016 January 22. Received 2016 January 21; in original form 2015 October 30}


\maketitle

\label{firstpage}

\begin{abstract}
Spectroscopic parameters (effective temperature, metallicity, etc) 
were determined for a large sample of $\sim 100$ red giants in the 
{\it Kepler} field, for which mass, radius, and evolutionary 
status had already been asteroseismologically established. 
These two kinds of spectroscopic and seismic information suffice to define 
the position on the ``luminosity versus effective temperature'' diagram 
and to assign an appropriate theoretical evolutionary track to each star. 
Making use of this advantage, we examined whether the stellar
location on this diagram really matches the assigned track,
which would make an interesting consistency check between theory 
and observation. It turned out that satisfactory agreement was confirmed 
in most cases ($\sim 90$\%, though appreciable discrepancies were
seen for some stars such as higher-mass red-clump giants), 
suggesting that recent stellar evolution calculations are practically reliable.
Since the relevant stellar age could also be obtained by this 
comparison, we derived the age--metallicity relation for these {\it Kepler} 
giants and found the following characteristics: (1) The resulting
distribution is quite similar to what was previously concluded 
for FGK dwarfs. (2) The dispersion of metallicity progressively
increases as the age becomes older. (3) Nevertheless, the maximum 
metallicity at any stellar age remains almost flat, which means
the existence of super/near-solar metallicity stars in a considerably
wide age range from $\sim$~(2--3)~$\times 10^{8}$~yr to $\sim 10^{10}$~yr.
\end{abstract}


\section{Introduction}

Thanks to the recent asteroseismological technique combined with very 
precise photometric continuous observations from satellites such as {\it Kepler}
or {\it CoRoT}, it has become possible to clearly discriminate the evolutionary 
status of red giants (shell H-burning phase before He-ignition or He-burning phase 
after He-ignition; cf. Bedding et al. 2011) and to accurately determine 
the stellar mass ($M$) as well as radius ($R$) by making use of the 
scaling relations (e.g., Pinsonneault et al. 2014, Casagrande et al. 2014,
and references therein).

Following this line, Takeda \& Tajitsu (2015; hereinafter referred to 
as Paper~I) conducted our first study based on high-dispersion spectra of
58 stars taken from Mosser et al.'s (2012) 218 sample of red giants in the {\it Kepler}
field with asteroseismologically established parameters, where they spectroscopically 
derived four atmospheric parameters for these 58 giant stars: effective temperature 
($T_{\rm eff}$), logarithmic surface gravity ($\log g$), microturbulent velocity 
($v_{\rm t}$), and metallicity ([Fe/H]; logarithmic Fe abundance relative to the Sun). 
Since their main purpose was to assess the accuracy of 
the stellar mass estimated from evolutionary tracks ($M_{\rm trk}$) 
as well as of the spectroscopic gravity ($\log g_{\rm spec}$) 
previously published by Takeda, Sato, \& Murata (2008) for a large number of field 
GK giants, they compared such conventionally determined parameters
of these {\it Kepler} sample with the corresponding seismic ones, and arrived
at the following conclusions: \\
--- (i) A satisfactory agreement was confirmed between $\log g_{\rm spec}$ and 
$\log g_{\rm seis}$, which may suggest that Takeda et al.'s (2008) gravity results 
are well reliable.\\
--- (ii) Meanwhile, Takeda et al.'s (2008) $M_{\rm trk}$ values for He-burning 
red-clump (RC) giants must have been considerably (typically by $\sim 50$\%) overestimated 
(presumably due to the ignorance of evolutionary status along with the use of 
incomplete set of evolutionary tracks), though those of H-burning red giants (RG) 
do not suffer such a problem.    

Now that the compatibility as well as reliability of seismic and spectroscopic
parameters has been confirmed, we can make use of them together in combination
with recent theoretically evolutionary tracks computed in very fine grids of
stellar parameters (cf. Appendix A in Paper~I).
This situation provides us with a good opportunity to examine the consistency
between the observed locations and theoretically computed evolutionary tracks of 
giant stars in the Hertzsprung--Russell (HR) diagram (i.e., $\log L$ vs. $\log T_{\rm eff}$ 
relation). That is, seismic $R_{\rm seis}$ and spectroscopic $T_{\rm eff,spec}$ 
suffice to define the location of each star on this diagram, while asteroseismologically
established stellar mass ($M_{\rm seis}$) and distinction of evolutionary status 
(RG or RC) [coupled with spectroscopically determined metallicity ([Fe/H]$_{\rm spec}$)] 
are sufficient to assign an appropriate evolutionary track to each star. 
``Does the observed location on the $\log L$--$\log T_{\rm eff}$ diagram
well matches the allocated evolutionary track?''
We thus decided to carry out this consistency check for a large number of {\it Kepler} 
giants, which have eventually added up to 106 stars (in combination with the previous 
58 stars in Paper~I) since we newly observed 48 stars for the present study.
This is the primary purpose of this paper.

An important by-product resulting from such comparison with theoretical tracks
is the age, which is mainly determined by the stellar mass in the present
case of giant stars (see, e.g., Casagrande et al. 2016). Thanks to the reliably 
known $M_{\rm seis}$, we can expect fairly precise age-evaluation for each star, 
by which the age--metallicity relation for these {\it Kepler} giant sample is 
finally accomplished, since the metallicity ([Fe/H]$_{\rm spec}$) is 
spectroscopically known. ``How can such established age--metallicity distribution 
for giants be compared with that derived for dwarfs?''
This examination is another aim of this investigation.

The remainder of this paper is organized as follows.
The new observational data of 48 {\it Kepler} giants (to be combined with the
previous 58 stars in Paper~I) and derivation of their spectroscopic parameters are
described in Sect.~3. We examine in Sect.~4 whether the location for each star 
on the HR diagram is consistent with the assigned evolutionary track.
The age--metallicity relation resulting from comparison with theoretical tracks 
is presented and discussed in Sect.~4, followed by Sect.~5 where the conclusions 
are summarised.

Besides, given that new data have been accumulated compared to the previous 
case in Paper~I and up-to-date theoretical tracks have become available,
we revisited the subjects treated in Paper~I and found some new enlightening
results, which are described in supplementary Appendices~A (comparison of 
spectroscopic and seismic $\log g$) and B (mass-determination 
from theoretical tracks).    

\section{Stellar parameters of new 48 Kepler giants}

\subsection{Observational data and spectroscopic parameter determination}

Our new spectroscopic observations for 48 giants in the {\it Kepler} field,
which were selected from Mosser et al.'s (2012) list, were carried out 
on 2015 July 3 (UT) by using Subaru/HDS and the data reduction was done 
by using IRAF in the same manner (i.e., with the same setting/procedure) 
as in Paper~I (cf. Sect.~2 therein for more details). The S/N ratios of
the resulting spectra (covering 5100--7800~$\rm\AA$) for these 48 stars 
turned out to be $\sim 100$ on the average, being similar to (or slightly 
worse than) the previous case of 42 stars in Paper~I. 

The atmospheric parameters ($T_{\rm eff}$, $\log g$, $v_{\rm t}$, and [Fe/H]) 
were determined by using the measured equivalent widths of Fe~{\sc i} and
Fe~{\sc ii} lines in the same way as in Paper~I (see Sect.~3.1 therein).
Also, the projected rotational velocity ($v_{\rm e}\sin i$) was evaluated by 
spectrum-fitting analysis applied to the 6080--6089~$\rm\AA$ region.
The final results for these 48 {\it Kepler} giants are summarised in Table~1, 
where the data are arranged in the same manner as in Table~1 of Paper~I.  
The equivalent-width data of Fe~{\sc i} and Fe~{\sc ii} lines along with 
the corresponding Fe abundances, and the detailed broadening/abundance results
of the 6080--6089~$\rm\AA$ fitting are also presented as supplementary 
online material (tableE1.dat and tableE2.dat). 
In analogy with Paper~I,  Fig.~1 (Fe abundance vs. equivalent width), 
Fig.~2 (Fe abundance vs. excitation potential), and Fig.~3 (spectrum-fitting
in 6080--6089~$\rm\AA$) are presented here (each corresponding to
Fig.~4, Fig.~5, and Fig.~6 of Paper~I, respectively).

\subsection{Special treatment for KIC~7341231}

Although the spectroscopic parameters of almost all newly observed stars were 
derived by exactly following the procedures of Paper~I as mentioned in Sect.~2.1, 
only one star (KIC~7341231 = BD~+42~3187) was exceptional. Actually, this is 
a very metal-poor ([Fe/H] = $-1.7$) subgiant of comparatively higher $T_{\rm eff}$ 
($\sim 5300$~K), belonging to the halo population characterized by considerably
large heliocentric radial velocity ($V_{\rm r}^{\rm hel} = -270$~km~s$^{-1}$).
Because of its conspicuously low metallicity along with higher $T_{\rm eff}$,
the metallic lines of this star are markedly weaker compared to other giants
(cf. Fig.~3), which makes it neither possible to determine the atmospheric parameters 
based on the adopted list of Fe~{\sc i}/Fe~{\sc ii} lines (Takeda et al. 2005),
nor to accomplish a reliable fitting analysis
in the 6080--6089~$\rm\AA$ region for $v_{\rm e}\sin i$ evaluation.
Accordingly, we employed a different set of stronger Fe~{\sc i}/Fe~{\sc ii} 
lines (see ``tableE1p.dat'' presented as online material), which were used by 
Takeda et al. (2006) for their study of RR~Lyr stars, in order to derive 
$T_{\rm eff}$, $\log g$, $v_{\rm t}$, and [Fe/H] of this star.
Similarly, its $v_{\rm e}\sin i$ derivation was done by fitting in the 
5200--5212~$\rm\AA$ region comprising strong lines of Cr~{\sc i}, Fe~{\sc i}, 
Ti~{\sc i} and Y~{\sc i} (see the inset in Fig.~3). 

\section{Comparison on the HR diagram}

\subsection{Location of each star}

Our total sample consists of 106 stars (42 stars from 2014 September observation
along with 16 stars from Thygesen et al.'s data as described in Paper~I, and
48 stars from 2015 July observation newly presented in this paper),
although the essential net number reduces to 103 because of the overlapping
of 3 stars in 2014 September and Thygesen et al. samples. Since all these stars
are taken from Mosser et al. (2012), their seismic radius ($R_{\rm seis}$) 
and mass ($M_{\rm seis}$) as well as the evolutionary stage (RG or RC1 or RC2)\footnote{
RG denotes red giants in the shell-H-burning phase before He ignition,
while RC indicates red-clump giants in the He-burning phase after He ignition, 
which are further classified into RC1 ($M<1.8$~M$_{\odot}$) and RC2 ($M>1.8$~M$_{\odot}$
according to the mass.} are already established.
In addition, spectroscopically determined effective temperature ($T_{\rm eff,spec}$)
and metallicity ([Fe/H]$_{\rm spec}$) are available from our study.
 
In this paper, we use the term ``HR diagram'' indicating a plot where stellar 
logarithmic effective temperature [$\log T_{\rm eff}$ ($\equiv X$), where $T_{\rm eff}$
is expressed in K] and logarithmic luminosity [$\log L/{\rm L}_{\odot}$ ($\equiv Y$)] 
are taken as the abscissa and ordinate, respectively.
We can naturally define the location of each star on this diagram, since $T_{\rm eff}$ 
is known and $L$ can be evaluated by the relation
\begin{equation}
\log (L/{\rm L}_{\odot}) = 4 \log (T_{\rm eff,spec}/{\rm T}_{{\rm eff},\odot}) 
                        +  2 \log (R_{\rm seis}/{\rm R}_{\odot}), 
\end{equation}
where quantities with $\odot$ are the reference solar values.
Such determined locations in the HR diagram for all the 106 stars are plotted 
(separated according to whether before or after He ignition) in Fig.~4a (RG) 
and Fig.~4b (RC1/RC2), where representative theoretical tracks corresponding to
$z = 0.01$ (see the next Sect.~3.2 for more details) are also drawn for comparison.

\subsection{Theoretical tracks}

Our next task is to assign an appropriate theoretical track to each star
in order to see whether it is consistent with the actual position.  
We use an extensive set of theoretical evolutionary tracks\footnote{
Available from {\tt http://stev.oapd.inaf.it/parsec\_v1.0/}
or {\tt http://people.sissa.it/\~{ }sbressan/parsec.html}}
computed by the Padova--Trieste group based on their PARSEC code (Bressan et al. 
2012, 2013). These tracks are provided with very fine grids in terms of $z$ 
(metallicity defined as the mass fraction of heavy elements) and $M$ (initial mass);
i.e.,  $z_{\rm grid}$ = {0.06, 0.04, 0.03, 0.02, 0.017, 0.014, 0.01, 0.008, 0.006, 
0.004, 0.002, 0.001, 0.0005, 0.0002, and 0.0001}, while the mesh of $M_{\rm grid}$ 
is 0.05~M$_{\odot}$ step (for 1--2.3~M$_{\odot}$) or 0.1~M$_{\odot}$ step 
(for 2.3--5~M$_{\odot}$). Since these parameter grids are sufficiently fine, 
we assign to each star the track corresponding to
($z_{\rm grid}$, $M_{\rm grid}$) being nearest to the actual 
($z_{\rm star}$, $M_{\rm star}$), where $M_{\rm star} \equiv M_{\rm seis}$
and $z_{\rm star} \equiv 0.014 \times 10^{\rm [Fe/H]_{spec}}$
(z$_{\odot} = 0.014$ is the solar value; cf. Asplund et al. 2009). 
The actual ($z_{\rm star}$, $M_{\rm star}$) as well as the adopted 
($z_{\rm grid}$, $M_{\rm grid}$) for each star are presented in Table~2.
Since the evolutionary stage is known for all the targets,  we naturally allocate 
``RG tracks'' (track portion from the point of ``end of core H-burning'' through 
the point of ``He ignition'') to RG-class stars, and ``RC tracks'' (track portion from 
the point of ``He ignition'' through the point of ``asymptotic giant branch tip'')\footnote{
Note that, in the PARSEC database, post-He-ignition tracks for 
$M \la 1.7$~M$_{\odot}$ (where He burning begins violently as ``He flash'' 
for this case of degenerated He core) are provided as independent data files 
labeled as ``HB'' (Horizontal Branch).} 
to RC1/RC2-class stars.

\subsection{Matching check}

Comparison of the position on the HR diagram with the assigned track for each star 
is graphically depicted in Fig.~5 (RG stars), Fig.~6 (RC1 stars), and Fig.~7 (RC2 stars).
We indicate in each panel the ``proximate point'' ($X^{*}_{\rm track}$, $Y^{*}_{\rm track}$) 
on the track by a Greek cross,
where the distance-measure $d^{2}(t)$ ($t$ is the age variable of a track) defined by
\begin{equation}
d^{2}(t) \equiv [(10 X_{\rm star} - 10 X_{\rm track}(t)]^{2} + [(Y_{\rm star} - Y_{\rm track}(t)]^{2}
\end{equation}  
becomes minimum.\footnote{
Here, an empirical weight factor of 10 is introduced for $X$ 
because of the practical reason to avoid inadequate solutions,
which sometimes result without it (especially for stars around 
the bottom of the ascending giant branch). While its choice 
is rather arbitrary, we found that it worked well with 10, 
which was chosen because the relevant span of 
$\log T_{\rm eff}$ is by $\sim 10$ times smaller than that of $\log L$
in the HR diagram under question.
}
These figures suggest that the agreement between ($X_{\rm star}$, $Y_{\rm star}$)
and ($X^{*}_{\rm track}$, $Y^{*}_{\rm track}$) is satisfactory in most cases, 
though appreciably discrepant cases sometimes show up (e.g., in RG and RC2 classes).
In order to clarify this situation quantitatively, the coordinate differences 
between the two points ($\Delta X \equiv  X^{*}_{\rm track} - X_{\rm star}$, 
$\Delta Y \equiv  Y^{*}_{\rm track} - Y_{\rm star}$) were computed (cf. Table~2)
and the $\Delta Y$ vs. $\Delta X$ plot is displayed in Fig.~8, 
from which the following consequences can be extracted.\\
--- Generally, a reasonable consistency to a level of $\Delta X \la 0.01$~dex 
(i.e., $\Delta T_{\rm eff} \la 100$~K) and $\Delta Y \la 0.1$~dex (i.e., $\la 0.2$~mag 
in magnitude) is accomplished for a majority ($\sim$~90\%) of our targets,
indicating reasonable reliability of recent stellar evolution calculations 
in the sense that theoretical tracks can satisfactorily reproduce 
the observed positions of red giant stars on the HR diagram.\\
--- However, appreciable discrepancies are sometimes seen especially in a group of 
RC2 stars (e.g., KIC 07581399, 07205067, 05307747, 05990753, 04902641, 09583430, 
09349632), the observed lumonosities of which are by $\sim$~0.2--0.4~dex 
lower than the theoretically predicted red-clump luminosities (cf. 
the corresponding panels in Fig.~7). A closer inspection in reference to Table~2
revealed that these stars have apparently higher mass values 
(2.5~${\rm M}_{\odot} \la M \la$~3.5~${\rm M}_{\odot}$) even among
RC2 stars (RC stars of $M > 1.8$~${\rm M}_{\odot}$). Actually, we can recognize 
from Fig.~4b that the luminosities of all RC2 stars (tiangles) tightly cluster 
at $\log (L/{\rm L}_{\odot}) \sim$~1.6--1.9 almost irrespective of their masses 
(1.8~${\rm M}_{\odot} \la M \la$~3.5~${\rm M}_{\odot}$), which apparently 
contradicts the theoretical RC2 tracks ($\log L$ increases by $\sim 0.5$~dex 
for a mass change from $\sim 2$~${\rm M}_{\odot}$ to $\sim 3$~${\rm M}_{\odot}$).
We can not find any reasonable explanation for this inconsistency
seen in higher-mass RC2 stars; it may be worthwhile to reexamine whether 
their assigned evolutionary tracks as well as $M$ and/or $R$ determination 
procedures are really valid.\\
--- Regarding RC1 and RG stars, the agreement is satisfactory in most cases, 
excepting that appreciable differences are seen in several stars;
e.g., KIC~05795626 for RC1, KIC~11717120, 07341231, and 08493735 for RG
(note that these RG stars are not so much on the ascending track of 
red-giants as rather subgiants). We also notice a tendency of weak (positive) 
correlation between $\Delta \log L$ and $\Delta \log T_{\rm eff}$ (cf. Fig.~8).

\section{Age--metallicity relation of giants}

\subsection{Derivation of age}

As a natural by-product of determining the proximate point on the track 
($X^{*}_{\rm track}$, $Y^{*}_{\rm track}$) described in Sect.~3.3, we 
can derive the stellar age as the corresponding $t$ value ($t^{*}$).
Note, however, that the age is mainly determined by the stellar mass 
in the present case of giant stars (see, e.g., Fig.~3c in Takeda et al. 2008), 
since it restricts the predominantly long lifetime on the main sequence, 
compared to which the period of post-main-sequence phase is insignificant. 
So, pinpointing the location on the giant track is not necessarily very 
important in this respect. In order to clarify this situation, the elapsed 
times at the track points of several critical evolutionary phases
as well as the corresponding fraction of main-sequence period are plotted 
against the stellar mass in Fig.~9, where we can see that these giants have 
spent a major fraction ($\sim$~60--90\%) of their past life on the main sequence. 

In order to maintain the consistency with the previous work,\footnote{
While zero-age main-sequence was usually adopted as the origin of age 
in many old calculations, computations are done from the pre-main sequence phase 
in the PARSEC tracks we adopted.} 
we define ``$age$'' as the time elapsed from the zero-age main sequence (ZAMS)
as $age \equiv t^{*} - t({\rm ZAMS})$. Such derived $age$ values are given in Table~2. 

\subsection{Result and implication}

The resulting $age$ vs. [Fe/H] distribution for our 106 {\it Kepler} giants 
is depicted in  Fig.~10a, and the comparison with the relation derived 
for field FGK dwarfs (Takeda 2007) is shown in Fig.~10b.
We can recognize from Fig.~10b that both results of giants and dwarfs
are quite consistent with each other, without any systematic discrepancy
such as that Takeda et al. (2008) once claimed (which is evidently due to 
erroneous overestimation of mass for many giants as pointed out in Paper~I). 

Regarding the observational study of galactic age--metallicity relation, 
not a few papers have been published so far (see, e.g., Sect.~1 of 
Bergemann et al. 2014 and the references therein). Although any consensus
has not yet been accomplished concerning the detailed characteristics,
it is no doubt that the metallicity at a given age is by no means single-valued
but more or less diversified. Our result implies that the [Fe/H] dispersion
tends to progressively increase with $age$ (from several $\times 10^{8}$~yr 
to $\sim 10^{10}$~yr) while the maximum metallicity does not change much
(i.e., near/super-solar level is retained over this large span of $age$), 
resulting in a ``right triangle-like'' shape.

Especially, the existence of very old ($age$ $\sim 10^{10}$~yr) metal-rich 
($0.0 \la$~[Fe/H]~$\la 0.4$) stars may be regarded as a significant consequence.
How should we interpret their origin? Were they born in the galactic bulge
and migrated to the present position over the long passage of time?
From this point of view, it would be interesting and worthwhile to study 
the chemical abundances of key elements (e.g., $\alpha$-group) of these old 
stars of high metallicity and to compare them with those of younger meta-rich stars.      

\section{Summary and conclusion}

Recent very high-precision photometric observations from satellites 
have enabled discrimination of the evolutionary status (RG/RC1/RC2) 
as well as determinations of $M$ and $R$ for a large number of 
red giant stars by exploiting the asteroseimological technique.  

Following the same manner in Paper~I where our first pilot study was done for 58 stars,
we determined in this study the atmospheric parameters ($T_{\rm eff}$, $\log g$, 
$v_{\rm t}$, and [Fe/H]) for additional 48 giants in the {\it Kepler} field 
by using Fe~{\sc i} and Fe~{\sc ii} lines.

Given that spectroscopic and seismic information is now available for these 
106 red giants in total, we could readily define the position on the 
$\log L$ vs. $\log T_{\rm eff}$ diagram and to assign an appropriate 
theoretical evolutionary track to each star. 

Our first aim was to examine whether the observed stellar location 
on this diagram really matches the assigned theoretical track.
We could confirm that the assigned track is mostly consistent with 
the actual position (to a level of $\la 100$~K in $T_{\rm eff}$ and 
$\la 0.2$~mag in $M_{\rm bol}$) for a majority ($\sim 90$\%) of our targets. 
Accordingly, we may state that recent stellar evolution calculations 
are reasonably reliable. 

However, appreciable inconsistencies are seen for $\sim 10$\%
of the sample stars. Especially noteworthy is that the luminosities of 
several RC2 stars with higher-$M$ (2.5~${\rm M}_{\odot} \la M \la$~3.5~${\rm M}_{\odot}$)
do not agree with the corresponding theorerical tracks, because they
tend to tightly cluster at $\log (L/{\rm L}_{\odot}) \sim$~1.6--1.9 
irrespective of their masses, which apparently contradicts the theoretical 
prediction. It may be worthwhile to reexamine the validity of assigned 
evolutionary tracks and of the $M$ as well as $R$ values for these stars.

Our second purpose was to establish the age--metallicity relation
based on these giant stars, since the stellar age could be as a natural 
by-product of location--track comparison on the HR diagram.
The resulting distribution for giants turned out to be in good agreement
with that for FGK dwarfs derived by Takeda (2007), which is characterized
by growing metallicity dispersion with an increase in age while
the maximum metallicity remains almost flat at the near/super-solar level
over the wide age range from $\sim$~(2--3)~$\times 10^{8}$~yr to $\sim 10^{10}$~yr.

The fact that very old ($age$ $\sim 10^{10}$~yr) metal-rich ($0.0 \la$~[Fe/H]~$\la 0.4$) 
stars do exist may be regarded as a significant consequence from the viewpoint
of galactic chemical evolution. Studying the chemical abundance characteristics
of these stars in detail would be worthwhile toward clarifying their origin. 

\section*{Acknowledgments}

This research has been carried our by using the SIMBAD database,
operated by CDS, Strasbourg, France.

\setcounter{table}{0}
\begin{table*}
\begin{minipage}{180mm}
\scriptsize
\caption{Basic data and the resulting parameters of the newly observed 48 stars.}
\begin{center}
\begin{tabular}{crccccrrrrcccc}\hline
KIC\# & $Kepler$ & $T_{\rm eff}$ & $\log g$ & $v_{\rm t}$ & [Fe/H] & $\nu_{\rm max}$ &
$\Delta \nu$ & $\Delta\Pi_{1}$ & $R_{\rm seis}$ & $M_{\rm seis}$ & $\log g_{\rm seis}$ & 
$v_{\rm e}\sin i$ & class \\
(1) & (2) & (3) & (4) & (5) & (6) & (7) & (8) & (9) & (10) & (11) & (12) &
(13) & (14) \\
\hline
02573092& 11.58& 4689& 2.48& 1.38& +0.00&  35.9&  4.08&293.80& 11.60& 1.43&  2.47&  2.1&RC1 \\
02696732& 11.50& 4821& 2.90& 1.04& $-$0.13&  90.4&  8.39& 66.55&  7.01& 1.33&  2.87&  1.9& RG \\
02988638& 12.16& 4912& 2.67& 1.22& +0.06&  91.4&  7.42&178.12&  9.14& 2.31&  2.88&  2.2&RC2 \\
03098045& 11.83& 4820& 2.34& 1.29& $-$0.24&  33.3&  4.15&281.10& 10.54& 1.11&  2.44&  2.1&RC1 \\
03323943& 11.61& 4826& 2.55& 1.26& $-$0.14&  31.0&  4.03&286.02& 10.41& 1.01&  2.41&  2.1&RC1 \\
03425476& 11.68& 4780& 2.57& 1.30& $-$0.03&  39.0&  4.42&303.60& 10.84& 1.37&  2.51&  2.2&RC1 \\
03531478& 11.64& 5000& 3.20& 0.98& $-$0.06& 243.5& 17.35& 87.60&  4.49& 1.50&  3.31&  1.8& RG \\
04039306& 11.55& 4806& 2.45& 1.30& $-$0.10&  32.5&  4.15&312.20& 10.27& 1.03&  2.43&  2.2&RC1 \\
04056266& 11.78& 5021& 2.67& 1.17& $-$0.03&  89.0&  7.38&270.80&  9.09& 2.25&  2.87&  2.1&RC2 \\
04350501& 11.74& 4929& 3.19& 1.01& $-$0.09& 139.0& 11.10& 69.30&  6.22& 1.63&  3.06&  2.0& RG \\
04448777& 11.56& 4805& 3.19& 0.95& +0.10& 220.4& 17.02& 89.90&  4.14& 1.13&  3.26&  1.5& RG \\
04570120& 11.64& 5035& 2.73& 1.20& +0.05&  90.8&  7.35&277.10&  9.37& 2.44&  2.88&  2.6&RC2 \\
04726049& 11.83& 5029& 3.25& 0.97& $-$0.16& 248.0& 18.03& 89.40&  4.25& 1.37&  3.32&  1.8& RG \\
05283798& 11.71& 4770& 2.53& 1.30& +0.09&  54.8&  5.33&268.98& 10.46& 1.79&  2.65&  2.2&RC2 \\
05514974& 11.36& 4674& 2.26& 1.30& $-$0.10&  32.2&  4.07&327.10& 10.44& 1.03&  2.42&  2.0&RC1 \\
05598645& 11.69& 5052& 3.44& 0.86& $-$0.17& 260.1& 19.59& 91.18&  3.78& 1.14&  3.34&  2.0& RG \\
05611192& 11.82& 5064& 2.91& 1.20& +0.02& 101.1&  7.99&216.70&  8.85& 2.43&  2.93&  3.0&RC2 \\
05858034& 11.74& 4887& 2.45& 1.28& $-$0.19&  37.6&  4.42&268.60& 10.56& 1.27&  2.49&  2.1&RC1 \\
06531928& 10.70& 5156& 3.73& 0.69& $-$0.57& 458.2& 31.58&115.80&  2.59& 0.95&  3.59&  2.0& RG \\
06579495& 11.86& 4772& 2.78& 1.03& $-$0.02&  85.6&  8.19& 73.80&  6.92& 1.22&  2.85&  1.2& RG \\
06665058& 11.49& 4750& 3.10& 0.86& $-$0.07& 107.6&  9.79& 77.40&  6.08& 1.18&  2.95&  2.0& RG \\
07341231&  9.91& 5305& 3.65& 1.31& $-$1.73& 368.0& 28.95&112.75&  2.51& 0.73&  3.50&  1.2& RG \\
07584122& 11.69& 4974& 3.29& 0.93& $-$0.08& 252.4& 18.57& 91.66&  4.05& 1.26&  3.33&  1.9& RG \\
07734065& 11.71& 4882& 2.20& 1.28& $-$0.43&  26.6&  3.68&318.40& 10.78& 0.93&  2.34&  1.9&RC1 \\
07799349&  9.47& 4969& 3.56& 0.94& +0.28& 562.7& 33.53&109.80&  2.77& 1.31&  3.67&  2.1& RG \\
08475025& 11.71& 4848& 2.88& 1.02& $-$0.06& 110.9&  9.66& 74.80&  6.50& 1.41&  2.96&  1.9& RG \\
08493735& 10.07& 5842& 3.62& 1.14& +0.01& 586.1& 38.86&113.12&  2.33& 1.05&  3.73&  3.9& RG \\
08751420&  6.91& 5260& 3.63& 0.95& $-$0.15& 536.4& 34.70&135.40&  2.54& 1.08&  3.66&  2.1& RG \\
09145955&  9.78& 4943& 2.85& 1.05& $-$0.34& 130.0& 11.00& 77.01&  5.93& 1.39&  3.04&  1.7& RG \\
09349632& 11.85& 4976& 2.75& 1.12& +0.12& 100.8&  7.79&226.80&  9.20& 2.60&  2.93&  2.6&RC2 \\
09583430& 11.60& 4854& 2.73& 1.19& +0.21& 102.2&  7.78&166.30&  9.24& 2.62&  2.93&  2.2&RC2 \\
09812421& 10.19& 5140& 3.48& 0.87& $-$0.21& 427.5& 27.85&112.00&  3.10& 1.27&  3.56&  2.0& RG \\
10382615& 11.68& 4890& 2.25& 1.27& $-$0.49&  31.6&  4.12&301.01& 10.22& 1.00&  2.42&  2.4&RC1 \\
10474071& 11.70& 4975& 2.65& 1.17& +0.09&  93.1&  7.50&266.80&  9.17& 2.38&  2.89&  2.1&RC2 \\
10600926& 11.64& 4879& 2.48& 1.29& $-$0.20&  27.7&  3.91&329.40&  9.94& 0.82&  2.36&  2.3&RC1 \\
10604460& 10.92& 4573& 2.37& 1.30& +0.14&  31.7&  3.80&308.30& 11.66& 1.26&  2.41&  3.6&RC1 \\
10709799& 10.82& 4522& 2.51& 1.11& $-$0.04&  36.9&  4.21& 57.20& 10.99& 1.29&  2.47&  1.7& RG \\
10866415& 11.02& 4791& 2.82& 0.93& $-$0.01&  94.4&  8.78& 67.70&  6.66& 1.25&  2.89&  2.0& RG \\
11177749& 10.90& 4677& 2.24& 1.26& $-$0.04&  33.6&  4.05&304.00& 11.00& 1.20&  2.44&  2.0&RC1 \\
11251115&  7.88& 4837& 2.57& 1.29& +0.07&  56.7&  5.08&295.20& 12.00& 2.45&  2.67&  2.2&RC2 \\
11352756& 10.96& 4604& 2.29& 1.31& $-$0.04&  27.0&  3.71&300.30& 10.45& 0.86&  2.34&  2.3&RC1 \\
11401156&  9.89& 5053& 3.58& 0.90& +0.10& 571.0& 35.78&115.00&  2.49& 1.09&  3.68&  1.8& RG \\
11618103&  7.70& 4922& 2.91& 1.10& $-$0.17& 106.0&  9.38& 74.40&  6.64& 1.41&  2.95&  1.6& RG \\
11717120&  9.27& 5087& 3.72& 0.75& $-$0.31& 623.2& 37.80&130.50&  2.44& 1.14&  3.72&  1.7& RG \\
11721438& 11.75& 4959& 2.87& 1.12& +0.14& 111.0&  8.53&177.60&  8.44& 2.40&  2.97&  2.4&RC2 \\
11802968& 10.89& 4962& 3.73& 0.75& $-$0.06& 498.9& 34.50&116.10&  2.32& 0.81&  3.62&  1.5& RG \\
12008680& 11.21& 4881& 2.55& 1.31& $-$0.32&  25.1&  3.65&321.80& 10.34& 0.81&  2.32&  2.7&RC1 \\
12070114& 10.86& 4698& 2.46& 1.30& +0.05&  41.2&  4.28&237.70& 12.10& 1.78&  2.53&  2.1&RC2 \\
\hline
\end{tabular}
\end{center}
The data of these new 48 stars are arranged in the same manner as in Table 1 of Paper~I.
Following the serial number and the $Kepler$ magnitude (in mag) of the {\it Kepler} Input Catalogue 
(KIC; cf. Brown et al. 2011) in Columns (1) and (2), the atmospheric parameters
(effective temperature $T_{\rm eff}$ in K, logarithmic surface gravity $\log g$ in 
cm~s$^{-2}$/dex, microturbulent velocity dispersion $v_{\rm t}$ in km~s$^{-1}$,
and metallicity [Fe/H] in dex) spectroscopically determined from Fe~{\sc i} and 
Fe~{\sc ii} lines  are presented in Columns (3)--(6).
Columns (7)--(9) give the asteroseismic quantities taken from 
Mosser et al. (2012): the central frequency of the oscillation power excess ($\nu_{\rm max}$
in $\mu$Hz), the large frequency separation ($\Delta\nu$ in $\mu$Hz), and the gravity-mode 
spacing ($\Delta\Pi_{1}$ in unit of s; good indicator for discriminating between RG and RC1/RC2). 
Presented in Columns (10)--(12) are the seismic radius (in R$_{\odot}$), seismic mass (in M$_{\odot}$), 
and the corresponding seismic surface gravity (in cm~s$^{-2}$/dex), which were evaluated 
from $\nu_{\rm max}$ and $\Delta\nu$ by using the scaling relations as in Paper~I.
In Columns (13) and (14) are given the projected rotational velocity (in km~s$^{-1}$;
derived from spectrum-fitting analysis) and the evolutionary class
determined by Mosser et al. (2012) (RG: red giant, RC1: 1st clump giant, RC2: 2nd clump giant).  
\end{minipage}
\end{table*}

\setcounter{table}{1}
\begin{table*}
\begin{minipage}{180mm}
\scriptsize
\caption{Matching results between stellar positions on the HR diagram and assigned evolutionary tracks.}
\begin{center}
\begin{tabular}{ccccccccccrc}\hline
star code& [Fe/H] & $z_{\rm star}$ & $z_{\rm grid}$ & $M_{\rm star}$ & $M_{\rm grid}$ & $X_{\rm star}$ & $Y_{\rm star}$ &
$\Delta X$ & $\Delta Y$ & $\log age$ & Figure \\ 
(1) & (2) & (3) & (4) & (5) & (6) & (7) & (8) & (9) & (10) & (11) & (12) \\
\hline
\multicolumn{12}{c}{[RG-class stars]}\\
t02696732R & $-$0.13& 0.0104& 0.0100& 1.33& 1.35& 3.6831& 1.376&+0.0037&+0.020&  9.521& Fig.5(1,1) \\
s03455760R & $-$0.07& 0.0119& 0.0100& 1.63& 1.65& 3.6678& 1.686&+0.0111&+0.070&  9.263& Fig.5(1,2) \\
t03531478R & $-$0.06& 0.0122& 0.0140& 1.50& 1.50& 3.6990& 1.053&+0.0022&+0.019&  9.429& Fig.5(1,3) \\
s03744043R & $-$0.35& 0.0063& 0.0060& 1.31& 1.30& 3.6943& 1.321&+0.0058&+0.028&  9.516& Fig.5(1,4) \\
n03744043R & $-$0.28& 0.0073& 0.0080& 1.31& 1.30& 3.6936& 1.317&+0.0006&+0.004&  9.545& Fig.5(1,5) \\
s04243623R & $-$0.31& 0.0069& 0.0060& 0.99& 1.00& 3.6994& 0.566&+0.0146&+0.094&  9.900& Fig.5(1,6) \\
t04350501R & $-$0.09& 0.0114& 0.0100& 1.63& 1.65& 3.6928& 1.311&+0.0089&+0.060&  9.260& Fig.5(2,1) \\
s04351319R & +0.29& 0.0273& 0.0300& 1.45& 1.45& 3.6881& 0.800&+0.0044&+0.031&  9.552& Fig.5(2,2) \\
t04448777R & +0.10& 0.0176& 0.0170& 1.13& 1.15& 3.6817& 0.913&+0.0040&+0.009&  9.830& Fig.5(2,3) \\
t04726049R & $-$0.16& 0.0097& 0.0100& 1.37& 1.35& 3.7015& 1.015&+0.0018&+0.006&  9.509& Fig.5(2,4) \\
s04952717R & +0.13& 0.0189& 0.0200& 1.24& 1.25& 3.6806& 0.987&+0.0026&+0.011&  9.713& Fig.5(2,5) \\
s05033245R & +0.11& 0.0180& 0.0170& 1.41& 1.40& 3.7032& 0.800&$-$0.0009&$-$0.008&  9.513& Fig.5(2,6) \\
s05530598R & +0.37& 0.0328& 0.0300& 1.68& 1.67& 3.6627& 1.340&+0.0108&+0.061&  9.359& Fig.5(3,1) \\
t05598645R & $-$0.17& 0.0095& 0.0100& 1.14& 1.15& 3.7035& 0.921&$-$0.0044&$-$0.016&  9.752& Fig.5(3,2) \\
s05723165R & $-$0.02& 0.0134& 0.0140& 1.36& 1.35& 3.7206& 0.710&$-$0.0005&+0.004&  9.541& Fig.5(3,3) \\
s05806522R & +0.12& 0.0185& 0.0170& 1.18& 1.20& 3.6603& 1.451&+0.0034&+0.018&  9.768& Fig.5(3,4) \\
s05866737R & $-$0.26& 0.0077& 0.0080& 1.52& 1.50& 3.6879& 1.577&+0.0014&+0.005&  9.363& Fig.5(3,5) \\
s06117517R & +0.28& 0.0267& 0.0300& 1.26& 1.25& 3.6674& 1.187&$-$0.0008&$-$0.003&  9.754& Fig.5(3,6) \\
s06144777R & +0.14& 0.0193& 0.0200& 1.18& 1.20& 3.6752& 1.153&+0.0000&+0.000&  9.745& Fig.5(4,1) \\
t06531928R & $-$0.57& 0.0038& 0.0040& 0.95& 0.95& 3.7123& 0.628&+0.0084&+0.059&  9.932& Fig.5(4,2) \\
t06579495R & $-$0.02& 0.0134& 0.0140& 1.22& 1.20& 3.6787& 1.347&$-$0.0033&$-$0.016&  9.743& Fig.5(4,3) \\
t06665058R & $-$0.07& 0.0119& 0.0100& 1.18& 1.20& 3.6767& 1.227&+0.0102&+0.045&  9.689& Fig.5(4,4) \\
n06690139R & $-$0.14& 0.0101& 0.0100& 1.56& 1.55& 3.6971& 1.396&$-$0.0010&$-$0.005&  9.343& Fig.5(4,5) \\
t07341231R & $-$1.73& 0.0003& 0.0002& 0.73& 0.75& 3.7247& 0.650&+0.0199&+0.127& 10.164& Fig.5(4,6) \\
t07584122R & $-$0.08& 0.0116& 0.0100& 1.26& 1.25& 3.6967& 0.954&+0.0042&+0.017&  9.622& Fig.5(5,1) \\
t07799349R & +0.28& 0.0267& 0.0300& 1.31& 1.30& 3.6963& 0.622&$-$0.0039&$-$0.027&  9.679& Fig.5(5,2) \\
t08475025R & $-$0.06& 0.0122& 0.0140& 1.41& 1.40& 3.6856& 1.320&$-$0.0012&$-$0.006&  9.519& Fig.5(5,3) \\
t08493735R & +0.01& 0.0143& 0.0140& 1.05& 1.05& 3.7666& 0.753&$-$0.0053&$-$0.299&  9.908& Fig.5(5,4) \\
s08702606R & $-$0.11& 0.0109& 0.0100& 1.09& 1.10& 3.7381& 0.636&+0.0016&$-$0.057&  9.803& Fig.5(5,5) \\
s08718745R & $-$0.25& 0.0079& 0.0080& 1.17& 1.15& 3.6904& 1.190&+0.0033&+0.011&  9.730& Fig.5(5,6) \\
t08751420R & $-$0.15& 0.0099& 0.0100& 1.08& 1.10& 3.7210& 0.646&$-$0.0011&$-$0.075&  9.808& Fig.5(6,1) \\
t09145955R & $-$0.34& 0.0064& 0.0060& 1.39& 1.40& 3.6940& 1.274&+0.0117&+0.055&  9.423& Fig.5(6,2) \\
t09812421R & $-$0.21& 0.0086& 0.0080& 1.27& 1.25& 3.7110& 0.779&+0.0036&+0.024&  9.580& Fig.5(6,3) \\
c10323222R & +0.04& 0.0154& 0.0140& 1.55& 1.55& 3.6556& 1.624&+0.0133&+0.082&  9.390& Fig.5(6,4) \\
t10709799R & $-$0.04& 0.0128& 0.0140& 1.29& 1.30& 3.6553& 1.656&+0.0051&+0.032&  9.630& Fig.5(6,5) \\
t10866415R & $-$0.01& 0.0137& 0.0140& 1.25& 1.25& 3.6804& 1.321&$-$0.0021&$-$0.011&  9.682& Fig.5(6,6) \\
t11401156R & +0.10& 0.0176& 0.0170& 1.09& 1.10& 3.7035& 0.559&$-$0.0046&$-$0.049&  9.888& Fig.5(7,1) \\
t11618103R & $-$0.17& 0.0095& 0.0100& 1.41& 1.40& 3.6921& 1.365&$-$0.0013&$-$0.006&  9.471& Fig.5(7,2) \\
t11717120R & $-$0.31& 0.0069& 0.0060& 1.14& 1.15& 3.7065& 0.553&+0.0152&+0.204&  9.675& Fig.5(7,3) \\
t11802968R & $-$0.06& 0.0122& 0.0140& 0.81& 0.80& 3.6957& 0.466&$-$0.0151&$-$0.015& 10.368& Fig.5(7,4) \\
\hline
\multicolumn{12}{c}{[RC1-class stars]}\\
s01726211C & $-$0.57& 0.0038& 0.0040& 1.19& 1.20& 3.6975& 1.873&+0.0028&+0.033&  9.608& Fig.6(1,1) \\
n01726211C & $-$0.57& 0.0038& 0.0040& 1.17& 1.15& 3.6931& 1.851&+0.0052&+0.057&  9.670& Fig.6(1,2) \\
s02303367C & +0.06& 0.0161& 0.0170& 1.23& 1.25& 3.6629& 1.695&+0.0052&+0.037&  9.712& Fig.6(1,3) \\
s02424934C & $-$0.18& 0.0092& 0.0100& 1.36& 1.35& 3.6805& 1.813&+0.0019&+0.017&  9.547& Fig.6(1,4) \\
t02573092C & +0.00& 0.0140& 0.0140& 1.43& 1.45& 3.6711& 1.766&+0.0060&$-$0.009&  9.495& Fig.6(1,5) \\
s02714397C & $-$0.47& 0.0047& 0.0040& 1.12& 1.10& 3.6911& 1.766&+0.0105&+0.040&  9.729& Fig.6(1,6) \\
n02714397C & $-$0.36& 0.0061& 0.0060& 1.14& 1.15& 3.6951& 1.787&+0.0000&+0.002&  9.708& Fig.6(2,1) \\
t03098045C & $-$0.24& 0.0081& 0.0080& 1.11& 1.10& 3.6830& 1.730&+0.0023&+0.019&  9.811& Fig.6(2,2) \\
s03217051C & +0.21& 0.0227& 0.0200& 1.22& 1.20& 3.6618& 1.662&+0.0025&+0.040&  9.758& Fig.6(2,3) \\
t03323943C & $-$0.14& 0.0101& 0.0100& 1.01& 1.00& 3.6836& 1.722&$-$0.0008&$-$0.005&  9.988& Fig.6(2,4) \\
t03425476C & $-$0.03& 0.0131& 0.0140& 1.37& 1.35& 3.6794& 1.740&$-$0.0010&+0.009&  9.589& Fig.6(2,5) \\
n03748691C & +0.11& 0.0180& 0.0170& 1.37& 1.35& 3.6778& 1.758&$-$0.0039&$-$0.020&  9.607& Fig.6(2,6) \\
s04036007C & $-$0.36& 0.0061& 0.0060& 1.38& 1.40& 3.6916& 1.798&+0.0048&+0.021&  9.441& Fig.6(3,1) \\
t04039306C & $-$0.10& 0.0111& 0.0100& 1.03& 1.05& 3.6818& 1.703&+0.0005&+0.024&  9.915& Fig.6(3,2) \\
s04044238C & +0.20& 0.0222& 0.0200& 1.06& 1.05& 3.6550& 1.616&+0.0063&+0.064& 10.012& Fig.6(3,3) \\
s04243796C & +0.11& 0.0180& 0.0170& 1.26& 1.25& 3.6646& 1.676&+0.0044&+0.043&  9.712& Fig.6(3,4) \\
s04445711C & $-$0.32& 0.0067& 0.0060& 1.35& 1.35& 3.6881& 1.789&+0.0075&+0.025&  9.484& Fig.6(3,5) \\
s04770846C & +0.02& 0.0147& 0.0140& 1.58& 1.60& 3.6855& 1.684&$-$0.0026&+0.047&  9.359& Fig.6(3,6) \\
s05000307C & $-$0.25& 0.0079& 0.0080& 1.41& 1.40& 3.7010& 1.794&$-$0.0083&+0.013&  9.470& Fig.6(4,1) \\
s05266416C & $-$0.09& 0.0114& 0.0100& 1.51& 1.50& 3.6782& 1.858&+0.0032&+0.029&  9.418& Fig.6(4,2) \\
t05514974C & $-$0.10& 0.0111& 0.0100& 1.03& 1.05& 3.6697& 1.668&+0.0086&+0.067&  9.913& Fig.6(4,3) \\
s05737655C & $-$0.63& 0.0033& 0.0040& 0.78& 0.80& 3.7012& 1.688&+0.0045&+0.032& 10.213& Fig.6(4,4) \\
n05795626C & $-$0.72& 0.0027& 0.0020& 1.21& 1.20& 3.6922& 1.751&+0.0253&+0.111&  9.549& Fig.6(4,5) \\
t05858034C & $-$0.19& 0.0090& 0.0100& 1.27& 1.25& 3.6890& 1.756&$-$0.0040&+0.011&  9.651& Fig.6(4,6) \\
t07734065C & $-$0.43& 0.0052& 0.0060& 0.93& 0.95& 3.6886& 1.772&+0.0022&$-$0.020& 10.001& Fig.6(5,1) \\
c08813946C & +0.09& 0.0172& 0.0170& 2.09& 2.10& 3.6868& 1.678&+0.0014&+0.004&  9.022& Fig.6(5,2) \\
t10382615C & $-$0.49& 0.0045& 0.0040& 1.00& 1.00& 3.6893& 1.728&+0.0118&+0.058&  9.873& Fig.6(5,3) \\
c10404994C & $-$0.06& 0.0122& 0.0140& 1.50& 1.50& 3.6815& 1.774&$-$0.0004&$-$0.002&  9.459& Fig.6(5,4) \\
n10426854C & $-$0.30& 0.0070& 0.0080& 1.78& 1.80& 3.6962& 1.892&$-$0.0034&$-$0.027&  9.218& Fig.6(5,5) \\
t10600926C & $-$0.20& 0.0088& 0.0080& 0.82& 0.80& 3.6883& 1.700&+0.0005&+0.006& 10.300& Fig.6(5,6) \\
\hline
\end{tabular}
\end{center}
\end{minipage}
\end{table*}

\setcounter{table}{1}
\begin{table*}
\begin{minipage}{180mm}
\scriptsize
\caption{(Continued.)}
\begin{center}
\begin{tabular}{ccccccccccrc}\hline
star code& [Fe/H] & $z_{\rm star}$ & $z_{\rm grid}$ & $M_{\rm star}$ & $M_{\rm grid}$ & $X_{\rm star}$ & $Y_{\rm star}$ &
$\Delta X$ & $\Delta Y$ & $\log age$ & Figure \\ 
(1) & (2) & (3) & (4) & (5) & (6) & (7) & (8) & (9) & (10) & (11) & (12) \\
\hline
\multicolumn{12}{c}{[RC1-class stars]}\\
t10604460C & +0.14& 0.0193& 0.0200& 1.26& 1.25& 3.6602& 1.726&+0.0039&$-$0.004&  9.731& Fig.6(6,1) \\
c10716853C & $-$0.08& 0.0116& 0.0100& 1.75& 1.75& 3.6879& 1.780&+0.0049&+0.047&  9.233& Fig.6(6,2) \\
t11177749C & $-$0.04& 0.0128& 0.0140& 1.20& 1.20& 3.6700& 1.715&+0.0023&+0.018&  9.753& Fig.6(6,3) \\
t11352756C & $-$0.04& 0.0128& 0.0140& 0.86& 0.85& 3.6631& 1.643&+0.0060&+0.026& 10.290& Fig.6(6,4) \\
n11444313C & +0.00& 0.0140& 0.0140& 1.39& 1.40& 3.6773& 1.797&$-$0.0005&$-$0.005&  9.547& Fig.6(6,5) \\
n11569659C & $-$0.26& 0.0077& 0.0080& 0.85& 0.85& 3.6883& 1.689&+0.0001&+0.003& 10.209& Fig.6(6,6) \\
n11657684C & $-$0.12& 0.0106& 0.0100& 1.27& 1.25& 3.6947& 1.824&$-$0.0097&$-$0.052&  9.652& Fig.6(7,1) \\
s11819760C & $-$0.18& 0.0092& 0.0100& 1.25& 1.25& 3.6834& 1.843&$-$0.0026&$-$0.026&  9.654& Fig.6(7,2) \\
t12008680C & $-$0.32& 0.0067& 0.0060& 0.81& 0.80& 3.6885& 1.735&+0.0029&+0.035& 10.264& Fig.6(7,3) \\
c12884274C & +0.11& 0.0180& 0.0170& 1.39& 1.40& 3.6705& 1.689&+0.0021&+0.038&  9.548& Fig.6(7,4) \\
\hline
\multicolumn{12}{c}{[RC2-class stars]}\\
s02013502S & $-$0.02& 0.0134& 0.0140& 1.94& 1.95& 3.6913& 1.740&+0.0013&+0.014&  9.187& Fig.7(1,1) \\
s02448225S & +0.16& 0.0202& 0.0200& 1.87& 1.85& 3.6606& 1.818&+0.0087&+0.067&  9.248& Fig.7(1,2) \\
t02988638S & +0.06& 0.0161& 0.0170& 2.31& 2.30& 3.6913& 1.639&+0.0048&+0.054&  8.911& Fig.7(1,3) \\
s03730953S & $-$0.07& 0.0119& 0.0100& 1.97& 1.95& 3.6867& 1.815&+0.0006&+0.008&  9.057& Fig.7(1,4) \\
s03758458S & +0.07& 0.0164& 0.0170& 2.18& 2.20& 3.6998& 1.792&$-$0.0045&$-$0.024&  9.038& Fig.7(1,5) \\
t04056266S & $-$0.03& 0.0131& 0.0140& 2.25& 2.25& 3.7008& 1.673&+0.0005&+0.004&  8.922& Fig.7(1,6) \\
t04570120S & +0.05& 0.0157& 0.0170& 2.44& 2.40& 3.7020& 1.704&$-$0.0041&+0.029&  8.860& Fig.7(2,1) \\
s04902641S & +0.03& 0.0150& 0.0140& 2.56& 2.60& 3.6978& 1.662&+0.0037&+0.214&  8.751& Fig.7(2,2) \\
s05088362S & +0.03& 0.0150& 0.0140& 2.22& 2.20& 3.6776& 1.933&+0.0005&+0.009&  8.946& Fig.7(2,3) \\
s05128171S & +0.04& 0.0154& 0.0140& 2.03& 2.05& 3.6820& 1.763&+0.0025&+0.017&  9.034& Fig.7(2,4) \\
t05283798S & +0.09& 0.0172& 0.0170& 1.79& 1.80& 3.6785& 1.705&+0.0055&$-$0.014&  9.226& Fig.7(2,5) \\
s05307747S & +0.01& 0.0143& 0.0140& 2.91& 3.00& 3.7017& 1.791&$-$0.0040&+0.339&  8.585& Fig.7(2,6) \\
t05611192S & +0.02& 0.0147& 0.0140& 2.43& 2.40& 3.7045& 1.664&$-$0.0023&+0.091&  8.845& Fig.7(3,1) \\
s05990753S & +0.19& 0.0217& 0.0200& 2.70& 2.80& 3.6999& 1.707&$-$0.0079&+0.241&  8.685& Fig.7(3,2) \\
s06276948S & +0.19& 0.0217& 0.0200& 2.35& 2.30& 3.6936& 1.655&+0.0002&+0.008&  8.926& Fig.7(3,3) \\
s07205067S & +0.03& 0.0150& 0.0140& 3.49& 3.40& 3.7045& 2.007&$-$0.0124&+0.369&  8.443& Fig.7(3,4) \\
s07581399S & +0.01& 0.0143& 0.0140& 3.13& 3.20& 3.7050& 1.850&$-$0.0100&+0.405&  8.511& Fig.7(3,5) \\
s08378462S & +0.06& 0.0161& 0.0170& 2.47& 2.40& 3.6985& 1.700&$-$0.0007&+0.033&  8.860& Fig.7(3,6) \\
s09173371S & +0.00& 0.0140& 0.0140& 2.32& 2.30& 3.7045& 1.653&$-$0.0025&+0.046&  8.896& Fig.7(4,1) \\
t09349632S & +0.12& 0.0185& 0.0170& 2.60& 2.60& 3.6969& 1.667&+0.0004&+0.183&  8.765& Fig.7(4,2) \\
t09583430S & +0.21& 0.0227& 0.0200& 2.62& 2.60& 3.6861& 1.628&+0.0074&+0.201&  8.775& Fig.7(4,3) \\
n09705687S & $-$0.19& 0.0090& 0.0100& 1.92& 1.90& 3.7100& 1.728&$-$0.0049&$-$0.028&  9.162& Fig.7(4,4) \\
t10474071S & +0.09& 0.0172& 0.0170& 2.38& 2.40& 3.6968& 1.664&+0.0011&+0.069&  8.860& Fig.7(4,5) \\
t11251115S & +0.07& 0.0164& 0.0170& 2.45& 2.40& 3.6846& 1.849&+0.0014&+0.009&  8.857& Fig.7(4,6) \\
t11721438S & +0.14& 0.0193& 0.0200& 2.40& 2.40& 3.6954& 1.587&$-$0.0013&+0.130&  8.874& Fig.7(5,1) \\
t12070114S & +0.05& 0.0157& 0.0170& 1.78& 1.80& 3.6719& 1.806&+0.0043&+0.041&  9.260& Fig.7(5,2) \\
\hline
\end{tabular}
\end{center}
Column (1) --- The first lower-case character denotes the data source (``s'' $\cdots$ 
our 2014 September observation, ``t'' $\cdots$ our 2015 July observation, ``n'' $\cdots$ 
Thygesen et al.'s NOT spectra, ``c'' $\cdots$ Thygesen et al.'s CFHT/TBL spectra).
followed by the KIC number (8 characters). The last upper-case character indicates
the evolutionary status (``R'' $\cdots$ RG, ``C'' $\cdots$ RC1, ``S'' $\cdots$ RC2).
Column (2) --- Observed logarithmic Fe abundance ratio relative to the Sun.
Column (3) --- Observed stellar metallicity (mass fraction of heavier elements) 
defined as $0.014 \times 10^{\rm [Fe/H]}$.  
Column (4) --- Metallicity of the evolutionary track assigned to each star. 
Column (5) --- Asteroseismologically evaluated stellar mass (in unit of 
${\rm M}_{\odot}$).
Column (6) --- Mass of the evolutionary track assigned to each star.
Column (7) --- Stellar $\log T_{\rm eff} [\equiv X_{\rm star}]$ (dex) where $T_{\rm eff}$ is in K.
Column (8) --- Stellar $\log L/{\rm L}_{\odot} [\equiv Y_{\rm star}]$ (dex). 
Column (9) --- Difference of ($X^{*}_{\rm track} - X_{\rm star}$), where $X^{*}_{\rm track}$ 
is the $X$ value of the proximate point on the assigned theoretical track closest to $(X_{\rm star}, Y_{\rm star})$.
Column (10) --- Difference of ($Y^{*}_{\rm track} - Y_{\rm star}$), where $Y^{*}_{\rm track}$ 
is the $Y$ value of the proximate point on the assigned theoretical track closest to $(X_{\rm star}, Y_{\rm star})$.
Column (11) --- Logarithmic stellar $age$ (dex) (measured from zero-age main-sequence) 
where $age$ is expressed in yr.
Column (12) --- Guide to the relevant figure panel, where Fig.$n(i,j)$ denotes
that the corresponding panel is at ($i$-th row, $j$-th column) of Fig.~$n$.
\end{minipage}
\end{table*}

\setcounter{figure}{0}
\begin{figure*}
\begin{minipage}{150mm}
\includegraphics[width=15.0cm]{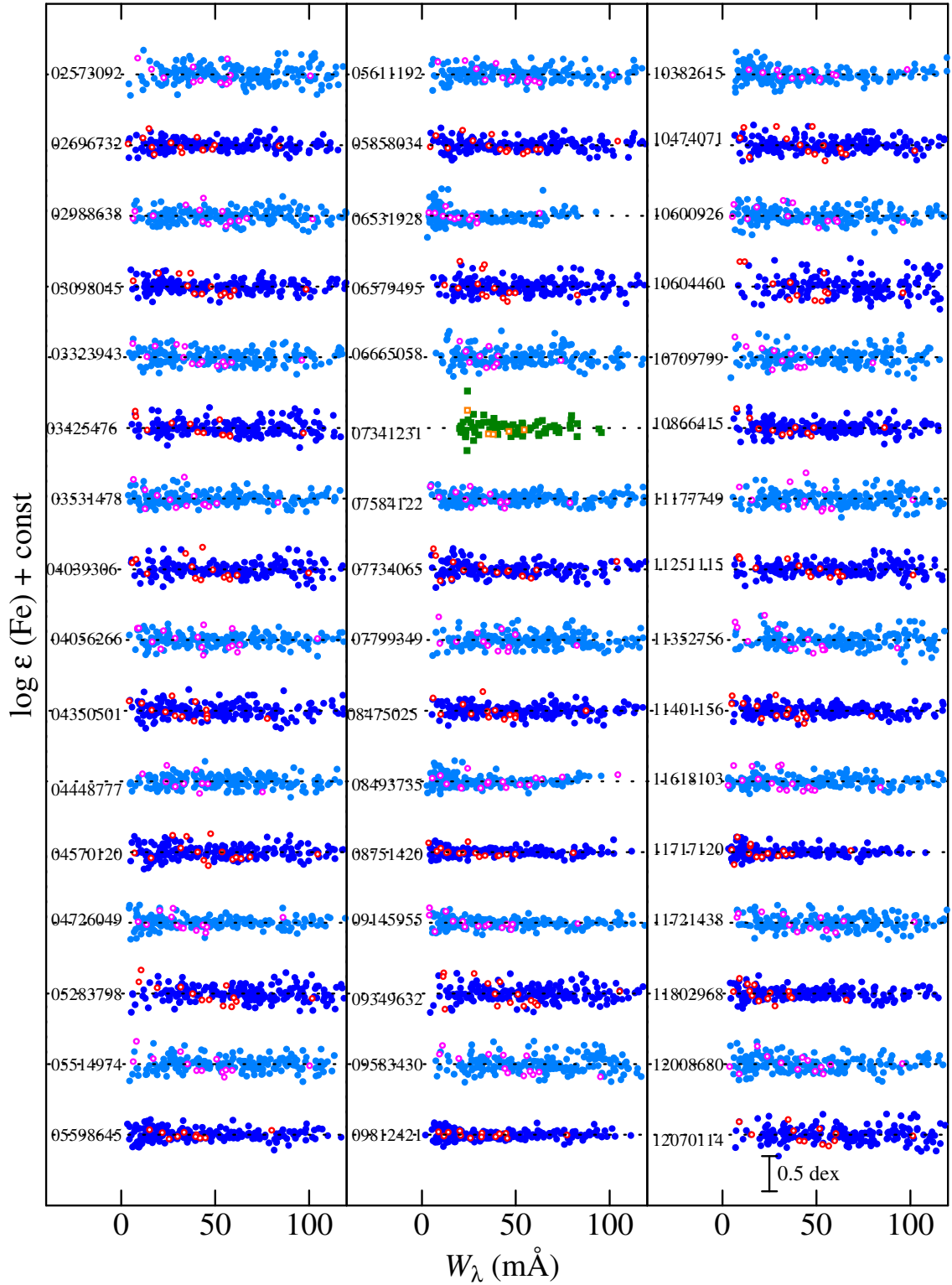}
\caption{Fe abundance vs. equivalent width relations 
corresponding to the finally established atmospheric parameters of 
$T_{\rm eff}$, $\log g$, and $v_{\rm t}$ for each of the new 48 stars,
being arranged in the increasing order of KIC number as in Table 1. 
The filled and open symbols correspond to Fe~{\sc i} and Fe~{\sc ii} 
lines, respectively. The results for each star are shown relative to 
the mean abundance (indicated by the horizontal dotted line), and 
vertically shifted by 1.0 relative to the adjacent ones.
Note that a different line-set was adopted for KIC~07341231
(very metal-poor subgiant) as explained in Sect.~2.2.
}
\label{fig1}
\end{minipage}
\end{figure*}

\setcounter{figure}{1}
\begin{figure*}
\begin{minipage}{150mm}
\includegraphics[width=15.0cm]{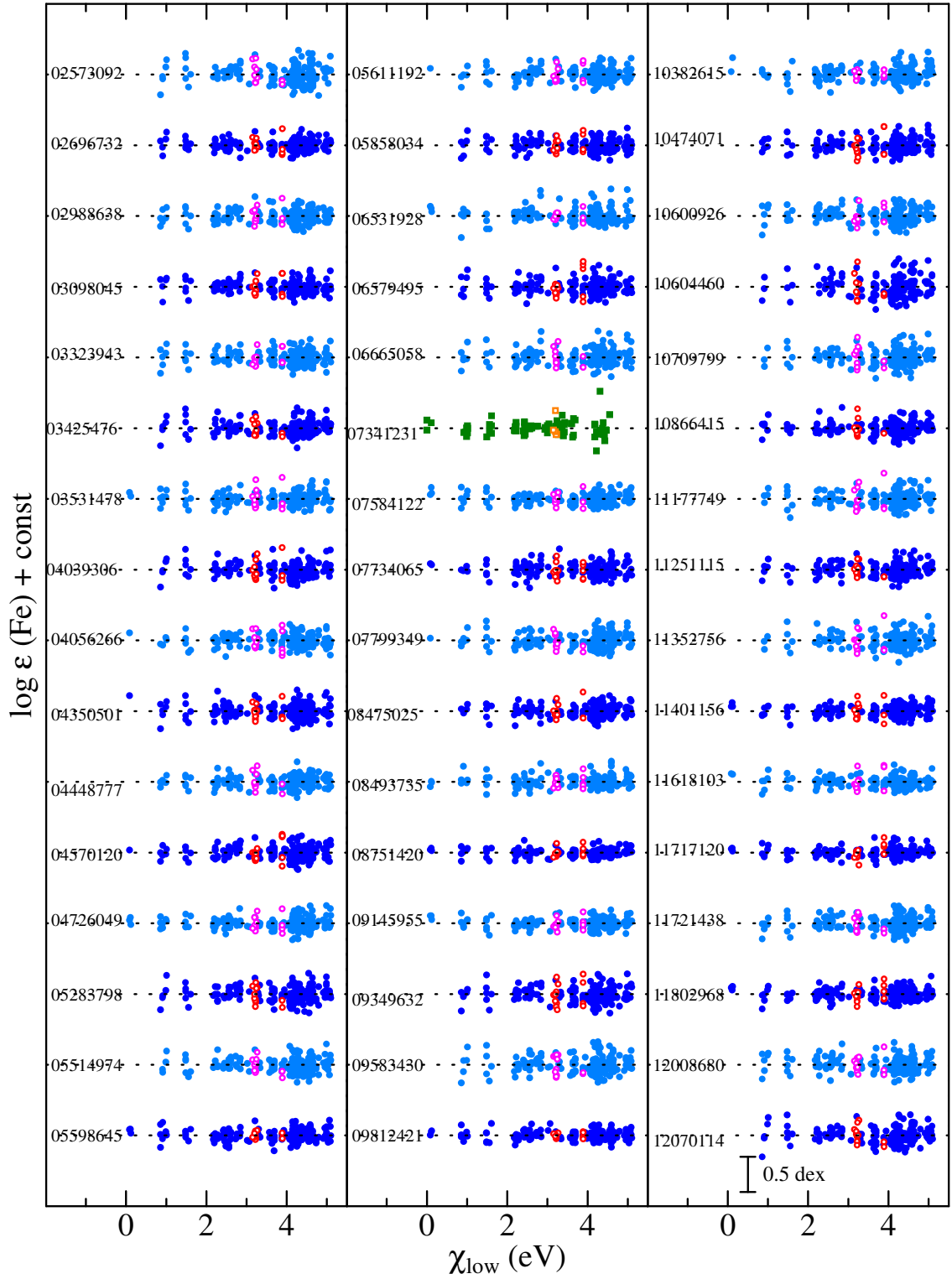}
\caption{Fe abundance vs. lower excitation potential relations 
corresponding to the finally established atmospheric parameters 
for each of the new 48 stars. Otherwise, the same as in Fig.~1.
}
\label{fig2}
\end{minipage}
\end{figure*}

\setcounter{figure}{2}
\begin{figure*}
\begin{minipage}{150mm}
\includegraphics[width=15.0cm]{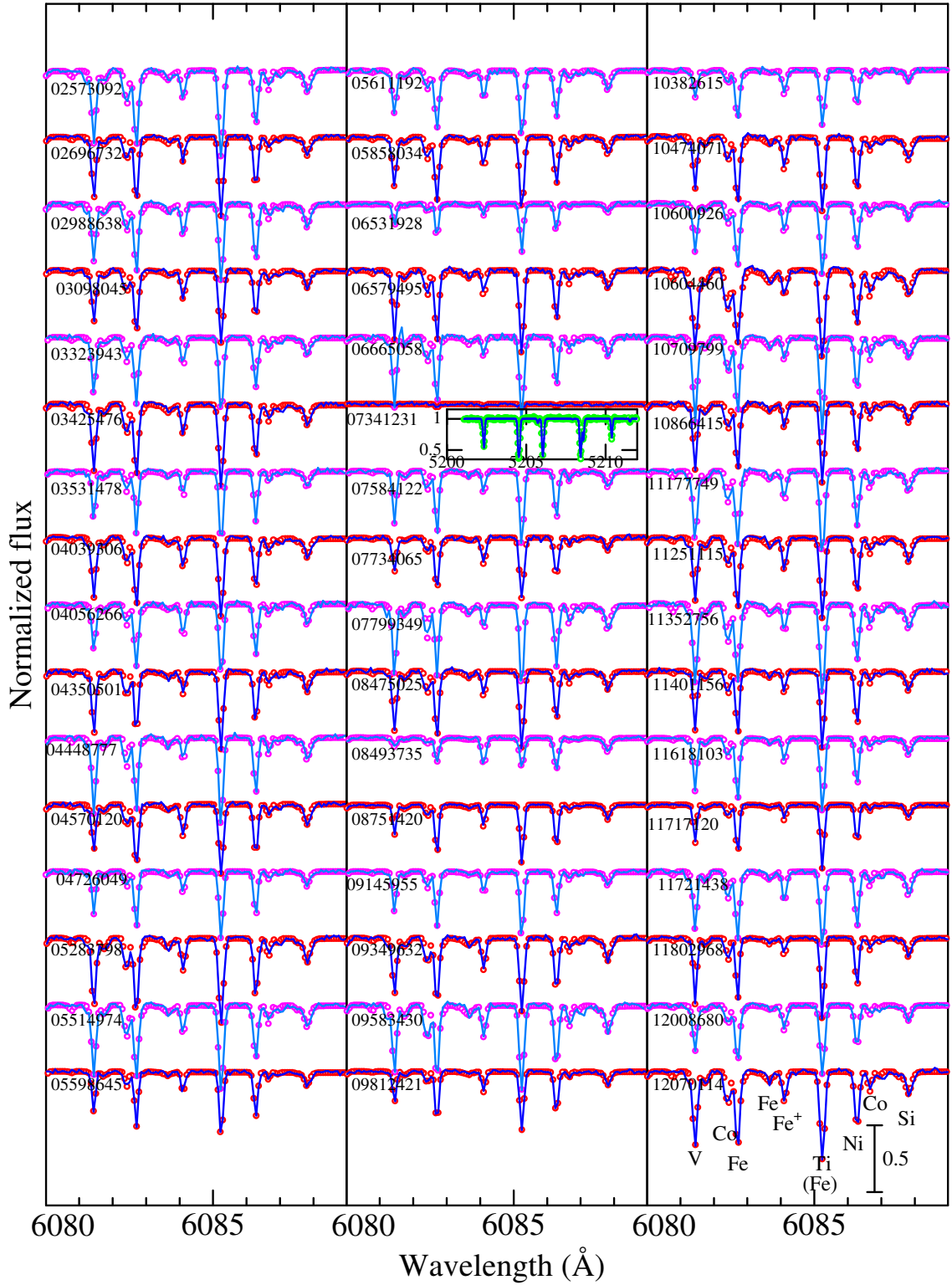}
\caption{Synthetic spectrum fitting in the 6080--6089~$\rm\AA$
region accomplished by varying the abundances of 
Si, Ti, V, Fe, Co, and Ni, along with the macrobroadening 
parameter and the wavelength shift (radial velocity).
The best-fit theoretical spectra are shown by solid lines, 
while the observed data are plotted by symbols, where
the wavelength scale of the stellar spectrum has been adjusted
to the laboratory frame. Each spectrum is vertically 
shifted by 0.5 relative to the adjacent one.
Note that spectrum fitting at the 5200--5212~$\rm\AA$ region
was specially applied to KIC~07341231 (as shown in the inset), 
since the lines in the 6080--6089~$\rm\AA$ region are too weak. 
The spectra are arranged in the same order as in Fig.~1.
}
\label{fig3}
\end{minipage}
\end{figure*}

\setcounter{figure}{3}
\begin{figure*}
\begin{minipage}{70mm}
\includegraphics[width=7.0cm]{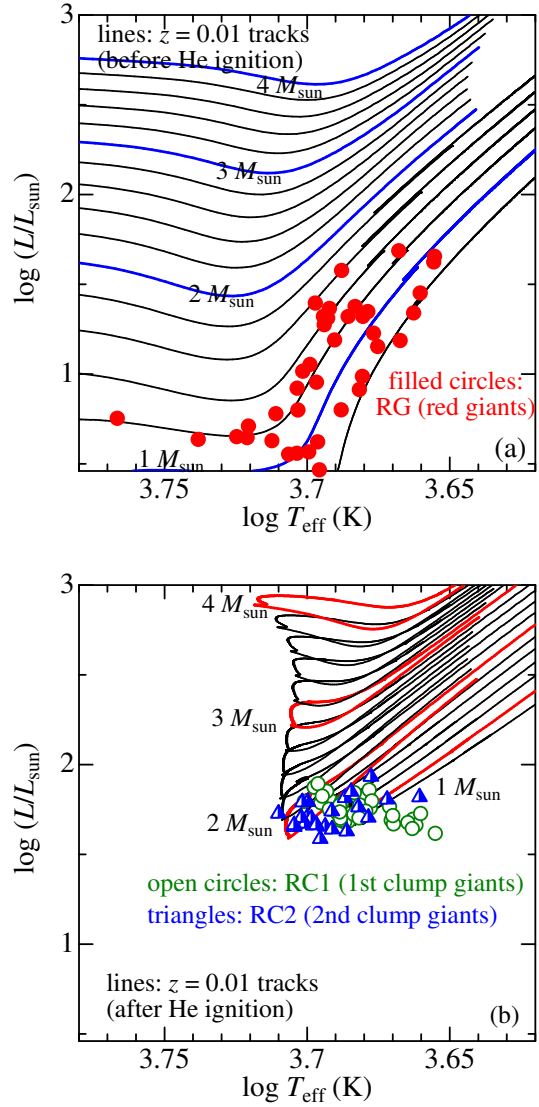}
\caption{
All the 106 program stars plotted on the $\log T_{\rm eff}$--$\log L/{\rm L}_{\odot}$
diagram. The upper panel (a) is for red giants in the shell-H-burning phase before He ignition (RG) 
while the lower panel (b) is for red-clump giants in the He-burning phase after He ignition 
(RC1 for $M<1.8$~M$_{\odot}$, RC2 for $M>1.8$~M$_{\odot}$).
Symbols are discriminated according to the evolutionary status:
Red filled circles $\cdots$ RG, green open circles $\cdots$ RC1, and blue triangles $\cdots$ RC2.
The PARSEC tracks corresponding to each evolutionary phase (either RG tracks or RC tracks) 
computed for $z = 0.01$ (slightly metal-deficient case by $\sim 0.2$~dex lower than
the solar metallicity) and various representative mass values are overplotted for comparison. 
}
\label{fig4}
\end{minipage}
\end{figure*}

\setcounter{figure}{4}
\begin{figure*}
\begin{minipage}{150mm}
\includegraphics[width=15.0cm]{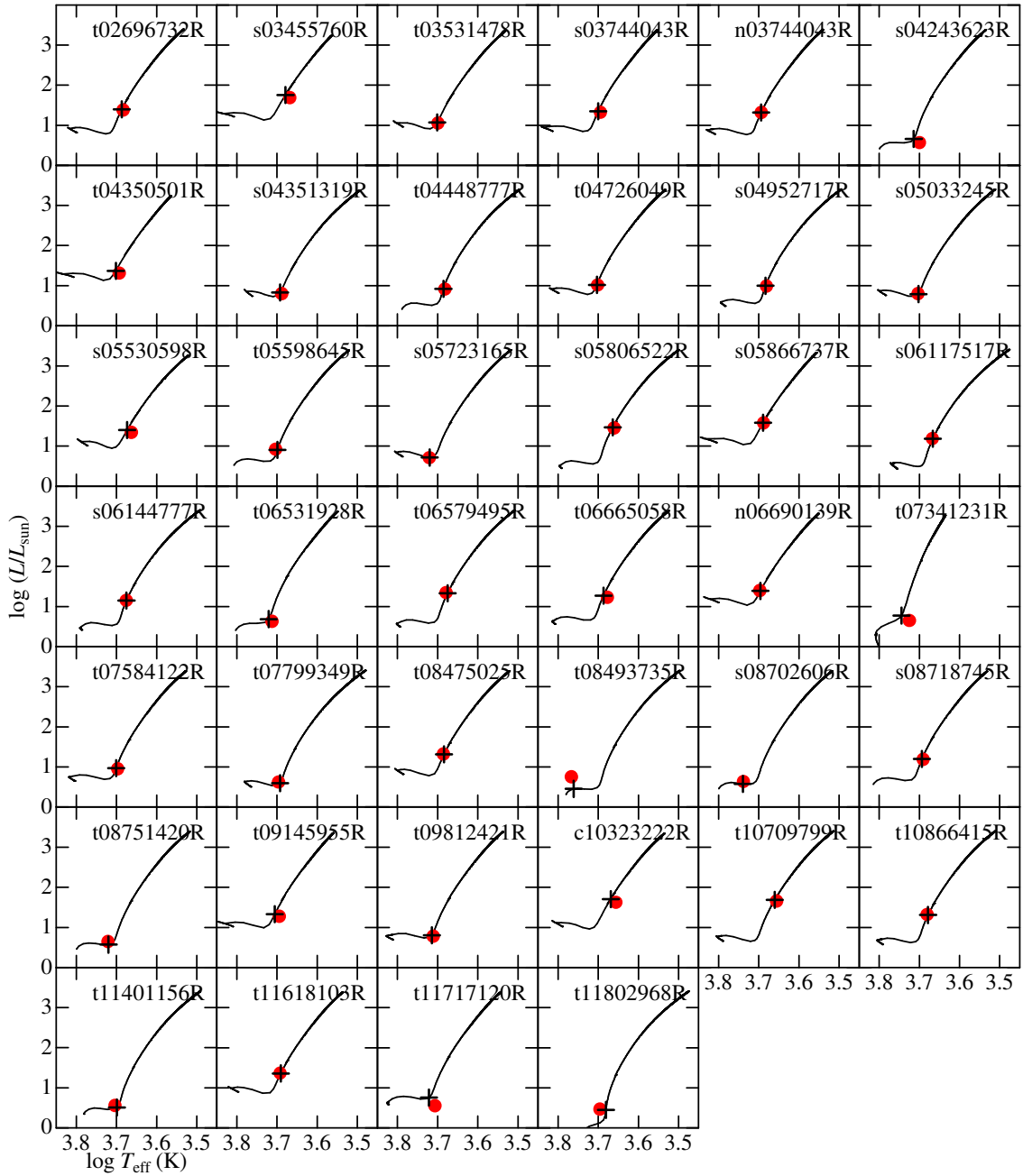}
\caption{
Comparison of the observed ($\log T_{\rm eff}$, $\log L/{\rm L}_{\odot}$)
with the assigned theoretical evolutionary track for 40 RG-class stars.
In each panel, the actual stellar location is indicated by a red filled circle,
while the corresponding proximate point on the track is expressed by a black Greek cross.
See Table~2 for the detailed data related to these figures.
}
\label{fig5}
\end{minipage}
\end{figure*}

\setcounter{figure}{5}
\begin{figure*}
\begin{minipage}{150mm}
\includegraphics[width=15.0cm]{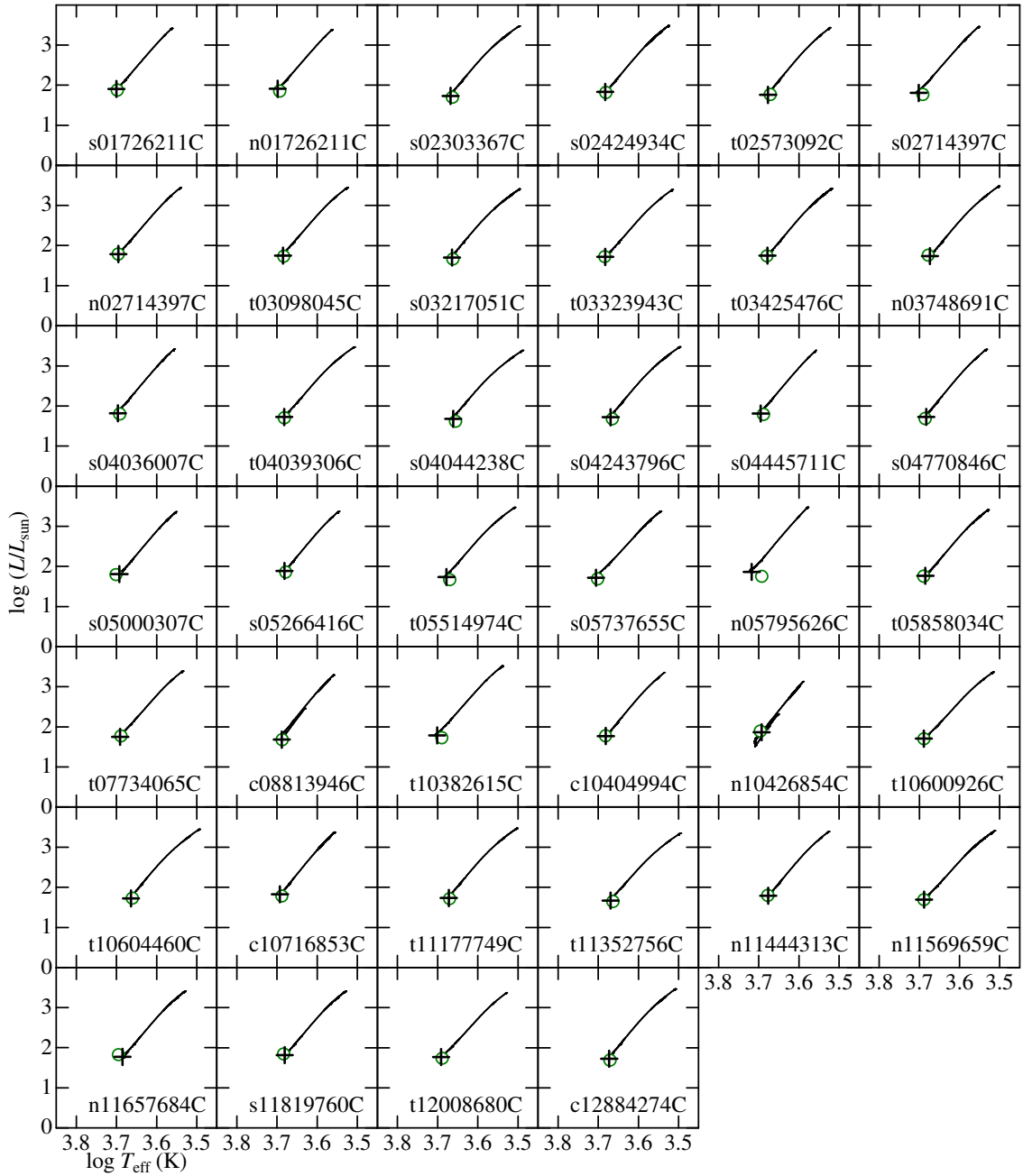}
\caption{
Comparison of the observed ($\log T_{\rm eff}$, $\log L/{\rm L}_{\odot}$)
with the assigned theoretical evolutionary track for 40 RC1-class stars.
In each panel, the actual stellar location is indicated by a green open circle.
Otherwise, the same as in Fig.~5.
}
\label{fig6}
\end{minipage}
\end{figure*}

\setcounter{figure}{6}
\begin{figure*}
\begin{minipage}{150mm}
\begin{center}
\includegraphics[width=15.0cm]{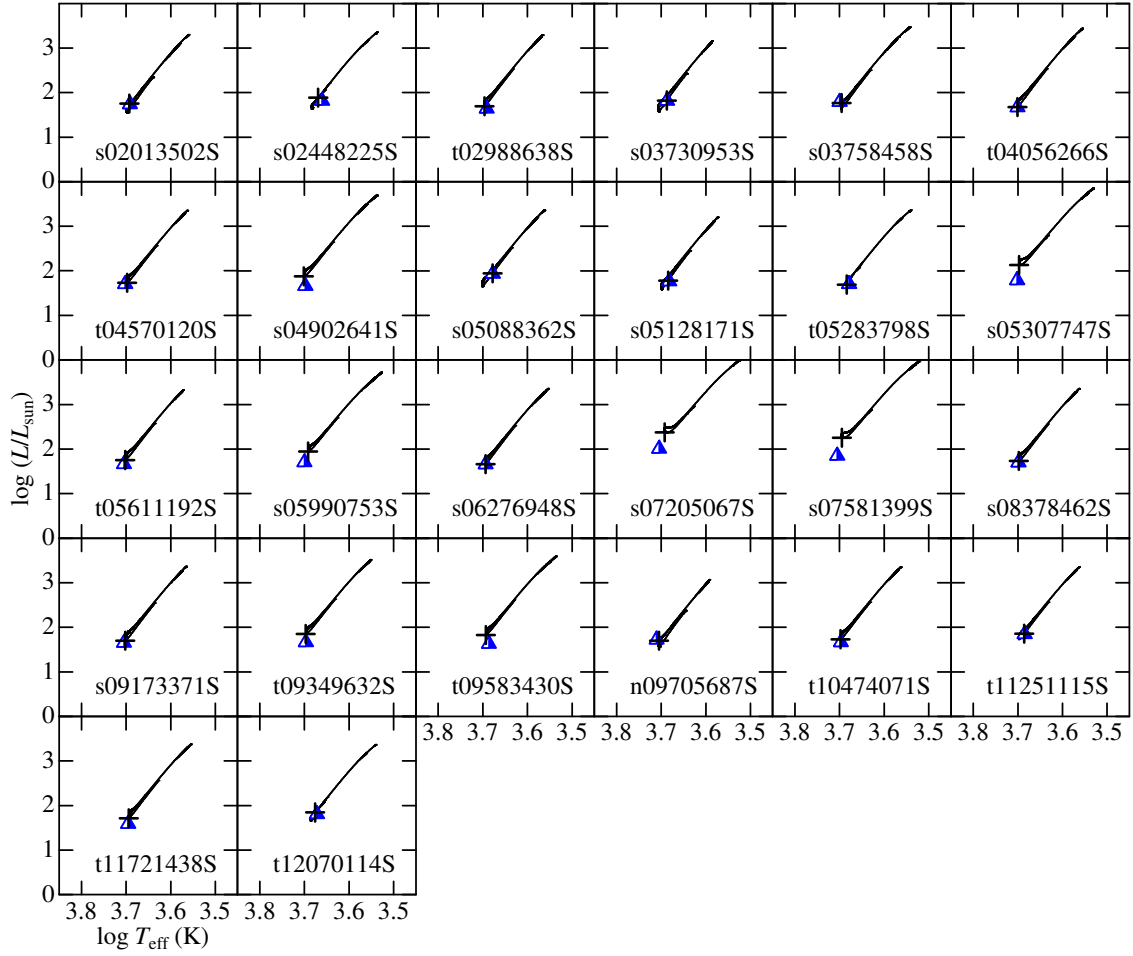}
\caption{
Comparison of the observed ($\log T_{\rm eff}$, $\log L/{\rm L}_{\odot}$)
with the assigned theoretical evolutionary track for 26 RC2-class stars.
In each panel, the actual stellar location is indicated by a blue triangle.
Otherwise, the same as in Fig.~5.
}
\label{fig7}
\end{center}
\end{minipage}
\end{figure*}

\setcounter{figure}{7}
\begin{figure*}
\begin{minipage}{80mm}
\begin{center}
\includegraphics[width=7.5cm]{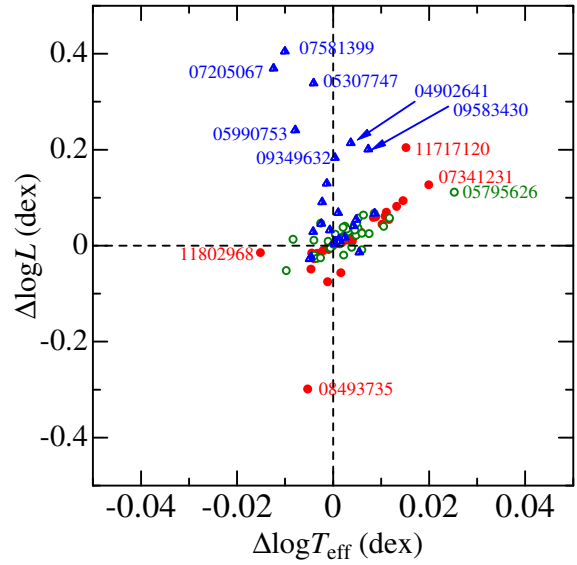}
\caption{
$\Delta \log L$ vs. $\Delta \log T_{\rm eff}$ (or $\Delta Y$ vs. $\Delta X$) 
plot based on the data given in Table~2, showing the behavior of 
theory$-$observation difference in the HR diagram for each star.
The meanings of the symbols are the same as in Fig.~4.
Stars showing rather large discrepancies are indicated in the figure. 
}
\label{fig8}
\end{center}
\end{minipage}
\end{figure*}

\setcounter{figure}{8}
\begin{figure*}
\begin{minipage}{120mm}
\begin{center}
\includegraphics[width=12cm]{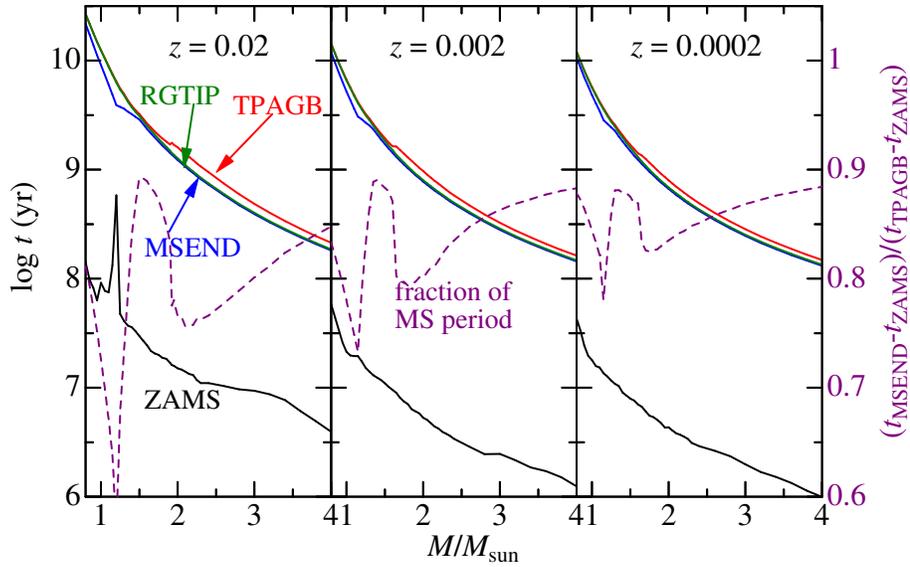}
\caption{
Elapsed times (measured from the beginning of star formation; expressed 
in logarithmic scale in unit of yr) of PARSEC stellar evolutionary
tracks at four critical phases are plotted against the stellar mass.
Black line, blue line, green line, and red line correspond to 
(i) zero-age main-sequence (ZAMS),  (ii) main-sequence end (MSEND), 
(iii) red-giant tip at He-ignition (RGTIP), and (iv) beginning of 
thermally-pulsing asymptotic giant-branch (TPAGB), respectively. 
The fraction of main-sequence period, which is defined as
$(t_{\rm MSEND} - t_{\rm ZAMS})/(t_{\rm TPAGB} - t_{\rm ZAMS})$, is also 
shown in purple dashed lines (its scale is given at the rightmost ordinate).
Each panel displays results for models of different metallicity:
$z = 0.02$ (left), $z= 0.002$ (middle), and $z=0.0002$ (right).
}
\label{fig9}
\end{center}
\end{minipage}
\end{figure*}

\setcounter{figure}{9}
\begin{figure*}
\begin{minipage}{75mm}
\begin{center}
\includegraphics[width=7.5cm]{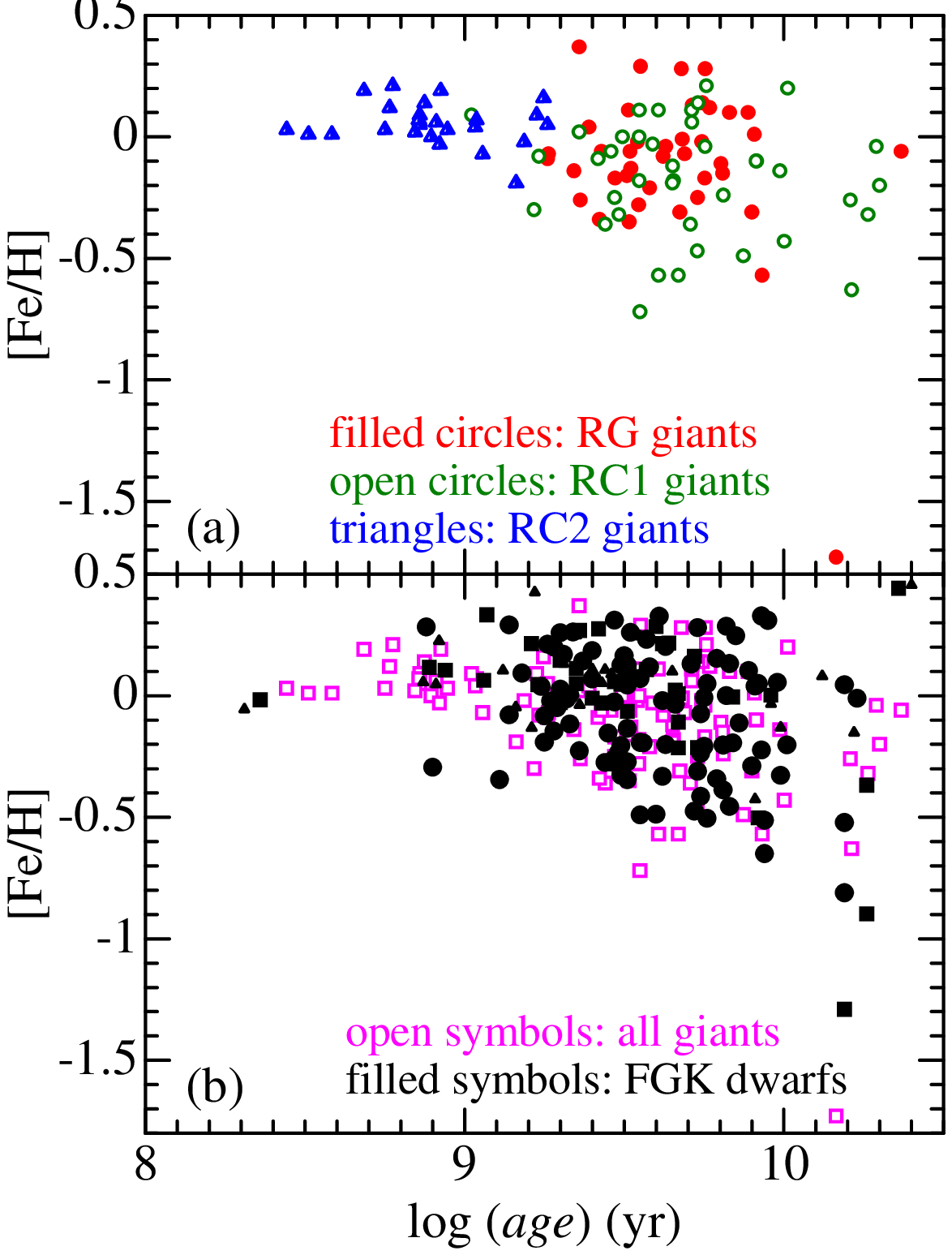}
\caption{
(a) Metallicity ([Fe/H]) vs. $age$ ($\equiv t - t_{\rm ZAMS}$) relation
derived for 106 {\it Kepler} giants, where stars for each evolutionary
phases are discriminated by symbols as in Fig.~4.
(b) Comparison of the [Fe/H] vs. $age$ relation of 106 {\it Kepler} giants
(pink open squares) with that of 160 FGK dwarfs (black filled symbols) 
derived by Takeda (2007). In the latter diagram, reliability classes
(see Sect. 2 of Takeda 2007) are discriminated by its type and size.
(A: reliable) $\cdots$ large circles, (B: less reliable) $\cdots$ 
medium-size squares, (C: unreliable) $\cdots$ small triangles.  
}
\label{fig10}
\end{center}
\end{minipage}
\end{figure*}

\appendix
\section{Metallicity-dependence in the scaling relation?}

One of the aims in Paper~I was to examine the consistency between the spectroscopic 
$\log g_{\rm spec}$ and the seismic $\log g_{\rm seis}$. Now that the sample size
has been almost doubled, it would be worthwhile to recheck this matter again.
The correlation between $\log g_{\rm spec}$ and $\log g_{\rm seis}$ is 
shown in Fig.~A1a, while the $\Delta\log g$(spec$-$seis) difference is plotted 
against $T_{\rm eff}$, $\log g$, and [Fe/H] in Fig.~A1b, A1c, and A1d, respectively.
We can confirm from these figures the same characteristics as concluded in Paper~I; 
i.e., (i) $\log g_{\rm spec}$ and $\log g_{\rm seis}$ satisfactorily agree with 
each other (with the standard deviation of $\sigma \sim 0.1$~dex) and (ii) 
$\Delta\log g$(spec$-$seis) difference does not show any systematic dependence 
upon $T_{\rm eff}$ and $\log g$.  

We also point out that this conclusion for $\Delta\log g$(spec$-$seis) holds
regardless of the evolutionary status (RG/RC1/RC2). Although Pinsonneault et al. 
(2014) recently carried out an extensive analysis on the stellar parameters of
a large number of red giants in the {\it Kepler} field and reported
a systematic disagreement in $\Delta\log g$(spec$-$seis) between RG and RC1/RC2
(i.e., negative for the former while positive for the latter; cf. Fig.~3
of their paper), such a trend can not be observed in our results. 

Interestingly, however, we newly noticed a slight metallicity-dependent trend in 
$\Delta\log g$(spec$-$seis), in the sense that $\Delta\log g$(spec$-$seis) 
tends to increase with a decrease in [Fe/H], as shown in Fig.~A1d.
It is hard to consider that such a [Fe/H]-dependence exists in spectroscopic 
gravities for the following reasons:\\
--- The difference in the metallicity for each star, which affects the opacity,
is properly taken into account in model atmospheres in our analysis.\\
--- It is unlikely that the non-LTE effect (non-LTE overionization of Fe~{\sc i})
is responsible, because it should lead to an {\it underestimation} of 
$\log g_{\rm spec}$ which becomes more conspicuous as the metallicity 
is lowered. That is, if this effect appreciably exists, $\Delta\log g$(spec$-$seis) 
would decrease with a decrease in [Fe/H], which is just the opposite to the trend 
seen in Fig.~A1d.

We, therefore, suspect that this effect may be attributed to $\log g_{\rm seis}$.
Here, the scaling relation for $\nu_{\rm max} (\propto g T_{\rm eff}^{-1/2})$ 
is relevant (while $\Delta \nu$ is irrelevant) in deriving $g_{\rm seis}$ 
(cf. Eq.(3) in Paper~I). However, this relation for $\nu_{\rm max}$, which was first 
proposed  by Kjeldsen \& Bedding (1995) in analogy with acoustic cut-off frequency,
is not so much physically justified as rather an empirically useful relation
(e.g., Bedding \& Kjeldsen 2003). Actually, Belkacem et al. (2013) discussed
based on their model calculation that this relation seems sufficiently good 
for dwarfs but some discrepancies (by up to $\sim 15$\%) may arise for giants. 
Accordingly, given that there is still room for further improvement for the expression
of $\nu_{\rm max}$, it may possibly depend upon the metallicity in some way. 

\setcounter{figure}{0}
\begin{figure*}
\begin{minipage}{120mm}
\begin{center}
\includegraphics[width=12.0cm]{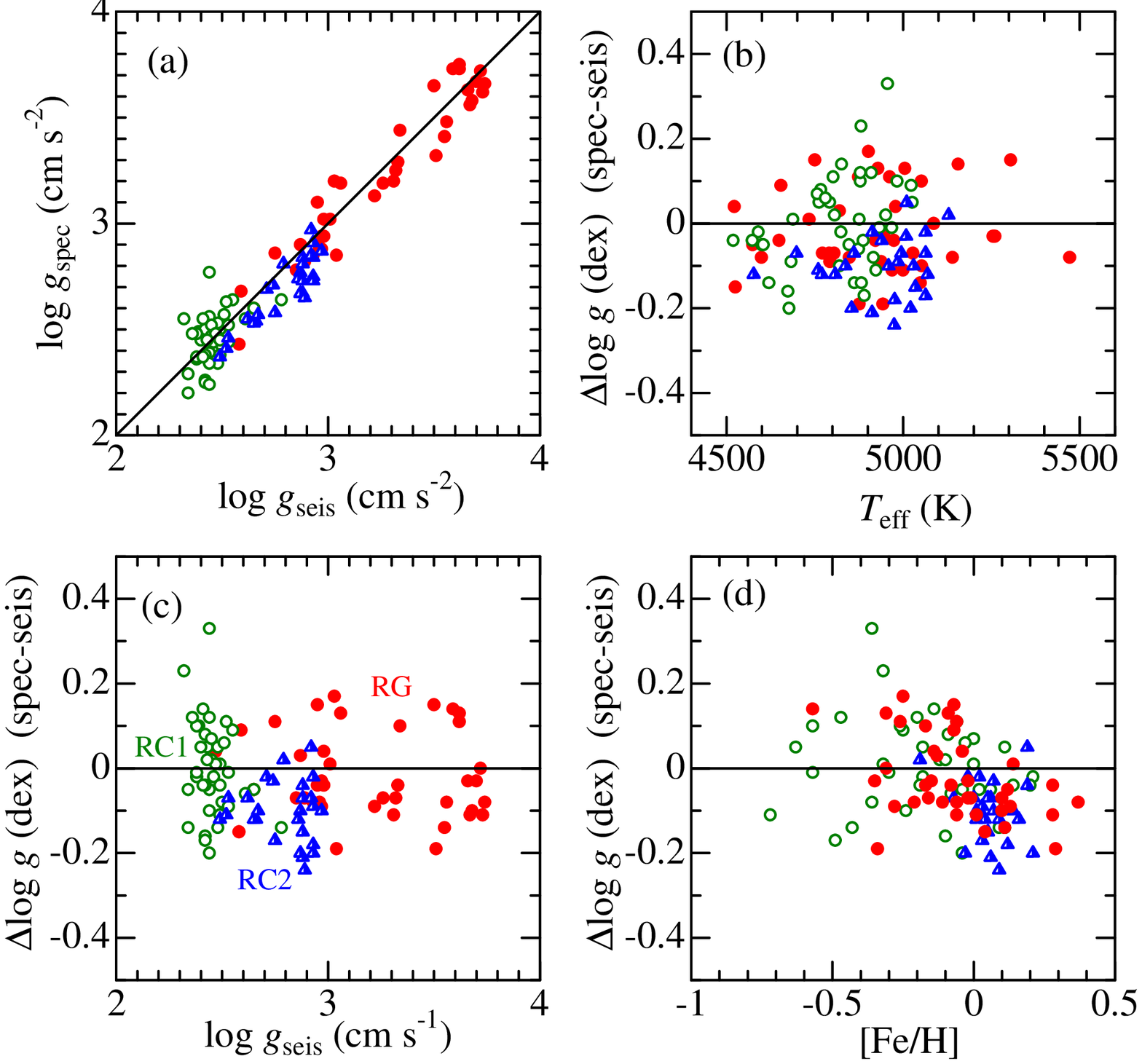}
\caption{
(a) Comparison between spectroscopic $\log g_{\rm spec}$ and 
seismic $\log g_{\rm seis}$ for 106 stars based the combined results 
of Paper~I and this study. (b) Plot of the $\log g$ difference 
[$\Delta \log g (\equiv \log g_{\rm spec} - \log g_{\rm seis})$] 
against $T_{\rm eff}$. (c) Plot of $\Delta \log g$ against $\log g_{\rm seis}$.
(d) $\Delta \log g$ plotted against [Fe/H].
The meanings of the symbols (discriminating the evolutionary
status of RG/RC1/RC2) are the same as in Fig. 4.
}
\label{figA1}
\end{center}
\end{minipage}
\end{figure*}

\section{Mass-determination problem revisited}

In Paper~I was studied how the stellar mass of a red giant star derived 
from evolutionary tracks ($M_{\rm trk}$) by following Takeda et al.'s (2008) 
procedure is compared with the seismic mass ($M_{\rm seis}$). 
They found that $M_{\rm trk}$ tends to be considerably overestimated 
(typically by $\sim 50$\% on the average) for RC-stars (cf. Fig.~12a therein),
which means that the mass values of many red clump giants in 
Takeda et al.'s (2008) sample must also be systematically too large.

Unfortunately, in Paper~I, the direct $M_{\rm seis}$ vs. $M_{\rm trk}$ was 
possible only for 9 {\it Kepler} giants (1 for RG, 6 for RC1, and 2 for RC2), 
because $L$ was determined from the apparent magnitude in the same manner 
as in Takeda et al. (2008) while the parallax data were available 
only for a limited number of stars.  
Since we have established this time the $L$ values for all 106 stars 
by combining $T_{\rm eff,spec}$ and $R_{\rm seis}$, we decided to
revisit this problem based on this large sample.   

In this test, we derived $M_{\rm trk}$ in three different ways:\\
--- (a) Exactly the same procedure as adopted by Takeda et al. (2008) 
was followed; i.e., combined RG+RC tracks of Lejeune \& Schaerer's (2001) grid 
for various mass values were used as if neither the mass nor 
the evolutionary status of each star were known (see Sect.~4.2 in Paper~I
for more details).\\
--- (b) Combined RG+RC tracks of the PARSEC grid (for various mass values 
but with the assigned $z_{\rm grid}$ closest to actual stellar metallicity) 
were used as if neither the mass nor the evolutionary status of each 
star were known. That is, $M_{\rm trk}$ was determined by searching for the minimum 
$d^{2}$ on the $(M,t)$ plane after trying all possible RG+RC tracks.\\
--- (c) Either RG or RC tracks of the PARSEC grid (for various mass values 
but with the assigned $z_{\rm grid}$ closest to actual stellar metallicity) 
were appropriately used depending on the known evolutionary status of each star.
That is, $M_{\rm trk}$ was determined by searching for the minimum 
$d^{2}$ on the $(M,t)$ plane after trying all possible RG tracks (for RG class)
or RC tracks (for RC1 or RC2 classes).

The resulting $M_{\rm trk}$ vs. $M_{\rm seis}$ plots corresponding to these
three cases are depicted in Fig.~B1a, Fig.~B1b, and Fig.~B1c, respectively.
We can see that Fig.~B1a is quite consistent with Fig.~12a in Paper~I,
indicating a considerable overestimation of $M_{\rm trk}$ for RC1 stars 
(red clump stars of lower mass), while the discrepancy is not
so large for RC2 stars (red clump stars of higher mass) where
$M_{\rm trk}$ tends to be even somewhat smaller than $M_{\rm seis}$
at the high-mass end ($M \ga 3 {\rm M}_{\odot}$).
We suspect that the reason why the considerably large overestimation 
of $M_{\rm trk}$ derived by Takeda et al.'s (2008) procedure is seen 
only in RC1 stars (but not in RC2 stars) is mainly due to the lack 
of ``He flash'' RC tracks for lower mass stars ($M <2$~M$_{\odot}$) 
in Lejeune \& Schaere's (2001) data (cf. Fig.~A1d in Paper~I), 
rather than the ignorance of the evolutionary status. 
Actually, we can recognize from the comparison of Fig.~B1b and Fig.~B1c
that the degree of consistency between $M_{\rm trk}$ and $M_{\rm seis}$
is nearly the same for both case (b) [combined RG+RC tracks were indifferently 
used] and case (c) [either RG or RC tracks were appropriately assigned]. 
This may suggest that the knowledge of the evolutionary status (RG or RC) 
in advance is not necessarily essential for deriving $M_{\rm trk}$ 
(in the sense that such information does not significantly improve 
the situation), for which sufficiently fine and wide coverage of 
the ($z$, $M$) grid in the adopted theoretical tracks (as well as
defining the stellar position on the HR diagram as precisely as possible) 
would be more important.  

\setcounter{figure}{0}
\begin{figure*}
\begin{minipage}{60mm}
\begin{center}
\includegraphics[width=6.0cm]{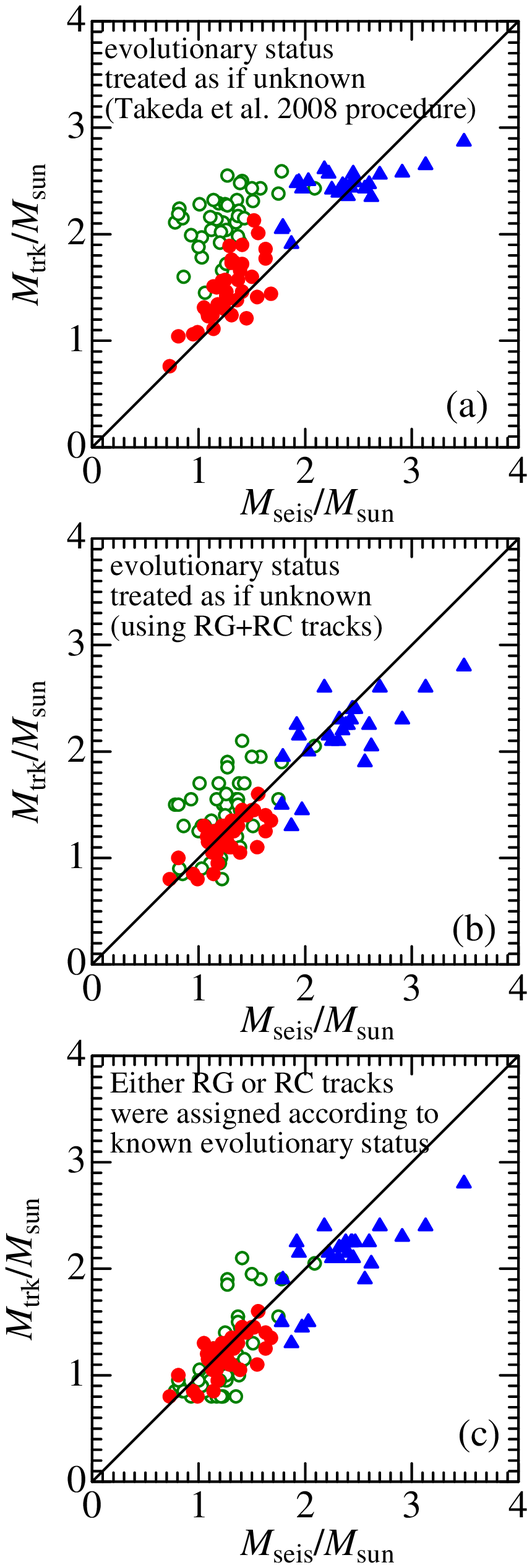}
\caption{
Comparison of asteroseismologically determined massses ($M_{\rm seis}$)
for 106 {\it Kepler} giants with those estimated from evolutionary tracks 
($M_{\rm trk}$), where three different kinds of tracks were tried. 
(a) Combined RG+RC tracks of Lejeune \& Schaerer's (2001) data
were used as if the evolutionary status of each star were unknown,
just as was done by Takeda et al. (2008).
(b) Combined RG+RC tracks of PARSEC data (for $z_{\rm grid}$
closest to actual stellar metallicity) were used as if 
the evolutionary status of each star were unknown.
(c) Either RG or RC tracks of PARSEC data (for $z_{\rm grid}$
closest to actual stellar metallicity) were appropriately
used depending on the known evolutionary status of each star.
See the caption of Fig.~4 for the meanings of the symbols.
}
\label{figB1}
\end{center}
\end{minipage}
\end{figure*}


\begin{thebibliography}{}
\bibitem[]{}
  Asplund M., Grevesse N., Sauval A. J., Scott P., 2009, ARA\&A, 47, 481
\bibitem[]{}
  Bedding T. R., et al., 2011, Nature, 471, 608
\bibitem[]{}
  Bedding T. R., Kjeldsen H., 2003, PASA, 20, 203
\bibitem[]{}
  Belkacem K., Samadi R., Mosser B., Goupil M.-J., Ludwig H.-G., 2013,
  in Progress in Physics of the Sun and Stars: A New Era in Helio- and Asteroseismology, 
  ASP Conf. Ser, Vol. 479, eds. H. Shibahashi \& A. E. Lynas-Gray, p. 61
  (San Francisco: Astronomical Society of the Pacific)
\bibitem[]{}
  Bergemann M., et al., 2014, A\&A, 565, A89
\bibitem[]{}
  Bressan A., Marigo P., Girardi L., Salasnich B., Dal Cero C., Rubele S., 
  Nanni A., 2012, MNRAS, 427, 127
\bibitem[]{}
  Bressan A., Marigo P., Girardi L., Nanni A., Rubele S., 2013, 
  EPJ Web of Conferences, 43, 3001 (DOI: http://dx.doi.org/10.1051/epjconf/20134303001)
\bibitem[]{}
  Brown T. M., Latham D. W., Everett M. E., Esquerdo G. A., 2011, AJ, 142, 112
\bibitem[]{}
  Casagrande L., et al., 2014, ApJ, 781, 110
\bibitem[]{}
  Casagrande L., et al., 2016, MNRAS, 455, 987
\bibitem[]{}
  Kjeldsen H., Bedding T. R., 1995, A\&A, 293, 87
\bibitem[]{}
  Lejeune T., Schaerer D., 2001, A\&A, 366, 538
\bibitem[]{}
  Mosser B. et al., 2012, A\&A, 540, A143
\bibitem[]{}
  Pinsonneault M. H., et al., 2014, ApJS, 215, 19
\bibitem[]{}
  Takeda Y., 2007, PASJ, 59, 335
\bibitem[]{}
  Takeda Y., Honda S., Aoki W., Takada-Hidai M., Zhao G., Chen Y.-Q., 
  Shi J.-R., 2006, PASJ, 58, 389
\bibitem[]{}
  Takeda Y., Ohkubo M., Sato B., Kambe E., Sadakane K., 2005, PASJ, 57, 27 (Erratum 57, 415)
\bibitem[]{}
  Takeda Y., Sato B., Murata D., 2008, PASJ, 60, 781
\bibitem[]{}
  Takeda Y., Tajitsu A., 2015, MNRAS, 450, 397 (Paper I)
\bibitem[]{}
  Thygesen A. O. et al., 2012, A\&A, 543, A160
\end{thebibliography}
\end{document}